\documentclass[a4paper,fleqn,usenatbib]{mnras}

\usepackage{graphicx}	

\def\aj{AJ}%
\def\actaa{Acta Astron.}%
\def\apj{ApJ}%
\def\apjl{ApJ}%
\def\apjs{ApJS}%
\def\ao{Appl.~Opt.}%
\def\aap{A\&A}%
\def\mnras{MNRAS}%
\def\pasj{PASJ}%
\title[Photometric metallicity map of the SMC]{Photometric metallicity map of the Small Magellanic Cloud}

\author[Choudhury S. et al.]{
Choudhury S.,$^{1}$\thanks{E-mail: samyaday.choudhury@gmail.com}
Subramaniam A.,$^{2}$
Cole A. A.,$^{3}$
and Sohn Y. -J.,$^{1,4}$
\\
$^{1}$Yonsei University Observatory, 120-749, Seoul, Republic of Korea\\
$^{2}$Indian Institute of Astrophysics, 2B Koramangala, Bangalore, India 560034\\
$^{3}$School of Natural Sciences, University of Tasmania, Private Bag 37, Hobart, Tasmania 7001, Australia\\
$^{4}$Department of Astronomy, Yonsei University, Seoul 03722, Republic of Korea
}

\date{Accepted XXX. Received YYY; in original form ZZZ}

\pubyear{2015}

\begin{document}
\label{firstpage}
\pagerange{\pageref{firstpage}--\pageref{lastpage}}
\maketitle

\begin{abstract}
We have created an estimated metallicity map of the Small Magellanic Cloud (SMC) using the Magellanic Cloud Photometric Survey (MCPS) and Optical Gravitational Lensing Experiment (OGLE III) photometric data. This is a first of its kind map of metallicity up to a radius of $\sim$ 2.5$^{\circ}$. We identify the RGB in the V, (V$-$I) colour magnitude diagrams of small subregions of varying sizes in both data sets. We use the slope of the RGB as an indicator of the average metallicity of a subregion, and calibrate the RGB slope to metallicity using available spectroscopic data for selected subregions. The average metallicity of the SMC is found to be [Fe/H] = $-$0.94 dex ($\sigma$[Fe/H] = 0.09) from OGLE III, and [Fe/H] = $-$0.95 dex ($\sigma$[Fe/H] = 0.08) from MCPS. We confirm a shallow but significant metallicity gradient within the inner SMC up to a radius of 2.5$^{\circ}$ ($-$0.045$\pm$0.004 dex deg$^{-1}$ to $-$0.067$\pm$0.006 dex deg$^{-1}$). 
\end{abstract}
\section{Introduction}

The Magellanic Clouds (MCs), comprising the Large Magellanic Cloud (LMC, d $\sim$ 50 kpc)and the Small Magellanic Cloud (SMC, d $\sim$ 60 kpc) are the closest pair of interacting galaxies to the Milky Way (MW). The LMC is a late-type spiral galaxy seen nearly face-on, whereas the structure of the SMC is not yet well-established. The gas and young stars are supposedly distributed in an highly-inclined disk of the SMC, whereas the old and intermediate-age stellar populations are found to be distributed in a regular, smooth spheroidal/ellipsoidal component \citep{Smitha&Purni2015}. This canonical vision of the structure of the SMC is being challenged by several recent works on variable stars, both old (RR Lyrae, e.g. \citealt{JD+2017AcA,Muraveva+2017arXiv}) and young (Classical Cepheids, e.g. \citealt{JD+2016AcA,Scowcroft+2016ApJ,Ripepi+2017MNRAS}). Optically, the appearance of the SMC is dominated by an elongated bar-like structure in the Northeast-Southwest direction, from which a prominent Wing extends to the east, in a direction that joins the Magellanic Bridge (MB) and the LMC \citep{Nidever+2011ApJ}. The relative proximity of these galaxies has allowed detailed analyses of the metallicity of individual field and cluster stars to probe their star formation and chemical enrichment history. 

The MCs are embedded within a common envelope of neutral hydrogen, indicating that these two galaxies are interacting with each other. It was also believed that the MCs have had interactions with the MW as well as amongst each other (\citealt{Murai&Fujimoto1980theMagStream, Tanaka1981theMagStream, Fujimoto&Murai1984theMagStream, Gardiner+1994numsimul, Westerlund1997theMCs}). However, recent measurements of their proper motion (Kallivayalil et al.2006a,b, 2013) suggest that they are approaching the MW for the first time (Besla et al. 2007, 2012). The MCs are metal poor (Z $\approx$ 0.008 for LMC and 0.004 for the SMC), gas rich, and have active ongoing star-formation, possibly triggered by interactions between themselves and/or interactions with the Galaxy. However, the mutual interaction between the Clouds, rather than the interaction with the MW, is fundamental in shaping their star formation history and metallicity gradients \citep{Cioni2009A&Athemetallicity}. 

The metallicity gradient (MG) within a galaxy can give indication of its formation and chemical evolution. This is especially interesting in the case of the SMC, which supposedly is dominated by its dynamical interaction history with the LMC and the MW. These interactions can trigger star formation, and produce changes in the chemical evolution of the galaxy over a certain period of time. During a collapse scenario gas is accreted and falls into the centre of the galaxy, where further star-formation takes place thus enriching the pre-existing gas \citep{Cioni2009A&Athemetallicity}. Stars may also form during the accretion process at large distances from the centre. The bar, disc, and halo components of a galaxy's potential can play a role in its chemical evolution. Also, dynamical interaction between galaxies and the accretion of satellites can alter the distribution of gas, modulating its chemical evolution and population gradients. Thus, accurate measurement of the MG is important to interpret the formation and evolution mechanisms.
 
The existence of a MG in the SMC has been a long standing debate. Previous studies have used star clusters (\citealt{Piatti+2007MNRASyoung, Piatti+2007MNRAS2newly, Parisi+2009AJclusI, Parisi+2015AJclusII}) as well as field stars (\citealt{Carrera+2008AJ-CEH-SMC, Cioni2009A&Athemetallicity, Piatti2012MNRASage-metalSMC, Dobbie+2014MNRAS-papII, Parisi+2016AJfieldII}) within the SMC in order to understand its MG. These studies included the use of spectroscopic as well as photometric data. 

\cite{Cioni2009A&Athemetallicity} used a photometric technique to estimate the MG of field asymptotic giant branch (AGB) stars within the SMC,  by comparing the ratio of carbon-rich (C-type) to oxygen-rich (M-type) AGBs. Using this C/M ratio as an indicator of [Fe/H] abundance, the authors estimated an almost constant metallicity ($-$1.25 $\pm$ 0.01 dex) to a galactocentric distance of about 12 kpc. Although these authors could cover a large area of the SMC, their indicators (AGB) and calibrators (Red Giant Branch stars) were different, and the C/M ratio is potentially susceptible to age effects. \cite{Parisi+2015AJclusII} used a sample of about 36 SMC star clusters by combining samples from \citep{Parisi+2009AJclusI} and other previous studies. They found a bimodality in the metallicity distribution of the SMC clusters with potential peaks at $-$1.1 and $-$0.8 dex but no strong MG. \cite{Piatti2012MNRASage-metalSMC} in an attempt to understand the age-metallicity relation in the SMC, analysed about 3.3 million field stars distributed throughout the entire main body using Washington photometric data. The author reported that the field stars do not possess gradients in age and metallicity. Also, the stellar populations formed since $\sim$2 Gyr ago are more metal rich than [Fe/H] $\sim$ $-$0.8 dex and are confined to the innermost region (semi-major axis $\leq$ 1$^{\circ}$ ). 

In work based on variable stars, \cite{Haschke+2012AJ} analyzed the Optical Gravitational Lensing Experiment, phase III (OGLE III) data of the SMC to obtain a mean metallicity of $-$1.70 $\pm$ 0.27 dex based on Fourier decomposition of I-band light curves of 1831 RR Lyraes. Their value was in very good agreement with earlier spectroscopic and photometric metallicities of RR Lyraes obtained by \cite{Butler+1982ApJ} and \cite{Kapakos+2011MNRAS} respectively. \cite{Haschke+2012AJ}, however, did not detect any MG for these older populations within the SMC. Also, \cite{Deb+2015MNRAS} did not detect any MG from the analysis of more than 1000 RR Lyrae stars using OGLE III data. Thus, all the above studies provided no evidence for any MG within the SMC.

On a different front there are studies that provide an evidence for a MG. \cite{Carrera+2008AJ-CEH-SMC} estimated a mean [Fe/H] $\sim$ $-$1.0 dex within the inner SMC, using CaT spectroscopy of over 350 red giant branch (RGB) stars in 13 fields distributed in different positions in the SMC (ranging from 1$^{\circ}$ to 4$^{\circ}$ from its center). The authors found that this mean metallicity decreases as one moves towards the outermost regions from centre. The most extensive spectroscopic study of RGB stars within the SMC has been carried out by \cite{Dobbie+2014MNRAS-papI,Dobbie+2014MNRAS-papII}. \cite{Dobbie+2014MNRAS-papII} used CaT spectroscopy of about 3000 stars within inner 5$^{\circ}$ of the SMC and confirmed a median [Fe/H] = $-$0.99 $\pm$ 0.01 with clear evidence for an abundance gradient of $-$0.075 $\pm$ 0.011 dex deg$^{-1}$.  Later on \cite{Parisi+2016AJfieldII} created an enlarged sample ($\sim$ 750) of RGB stars using their previous study \citep{Parisi+2010AJfieldI} and estimated a median metallicity of [Fe/H] $\sim$ $-$0.97$\pm$ 0.01, and detected a gradient of $-$0.08$\pm$0.02 dex deg$^{-1}$ within the inner 4$^{\circ}$. This value of MG was similar to that estimated by \cite{Dobbie+2014MNRAS-papII}. 

Rubele et al. (2015) attempted to understand the star formation history across the main body and Wing of the SMC (14 sq.\ degree) using near infrared data (VISTA survey of the MCs) by employing a colour magnitude diagram (CMD) reconstruction method. The authors analysed about 120 sub-regions, each covering  21$^{\prime}$ $\times$21.5$^{\prime}$, and presented the spatial distribution of mean metallicity for populations of three different ages ($\sim$ 800 Myr, 1.3 Gyr, and 2 Gyr). They found mean metallicity between $-$0.4 $\geq$ [Fe/H] $\geq$ $-$0.85 dex for each age bin. However, the authors do not estimate the MG within the SMC. \cite{Kapakos&Hatz2012MNRAS} performed Fourier decomposition analysis of 8- and 13-year V-band light curves of 454 fundamental-mode RR Lyrae variables from the OGLE III Catalogue of Variable Stars. The authors estimated an average metal abundance of $-$1.69 $\pm$ 0.41 dex, with a tentative MG ($-$0.013 $\pm$ 0.007 dex kpc$^{-1}$) that indicated increasing metal abundance towards the dynamical centre of the SMC.

In a previous attempt to understand the metallicity variation within the Large Magellanic Cloud, \cite{Choudhury+2016MNRAS} (hereafter Paper I) created a first of its kind, high-spatial resolution metallicity map with RGB stars as the tool, using the Magellanic Cloud Photometric Survey (MCPS) and OGLE III photometric data. The RGB is identified in the V, (V $-$ I) colour-magnitude diagrams (CMDs) of small sub-regions of varying sizes in both data sets. The slope of the RGB is used as an indicator of the mean metallicity of a sub-region, and it is calibrated to metallicity using spectroscopic data for field and cluster RGB stars in selected sub-regions. The study reconfirms that reliable photometric metallicity estimates can cover a large area of the galaxy, unlike the spectroscopic method, which can over only relatively small areas. The photometric method can bring out the overall distribution, variation, and global average of metallicity. It also can identify regions which might show large deviation with respect to the mean value. 

A high-spatial metallicity map showing the metallicity trend across the inner SMC is still unavailable. Also, it is seen that there has been less consensus over the nature of the MG within the SMC. Thus, we require a study using spatially extensive and homogeneous data sets to make significant advances. In this study, we extend the techniques developed in Paper I to the SMC with similar data sets to estimate a first of its kind high-spatial resolution metallicity map for the smaller Cloud. 

The paper is organised in the following way: In Section 2 we describe the data (OGLE III and MCPS) used in this study. The OGLE III analysis, metallicity maps and results are presented in Section 3, whereas those corresponding to the MCPS data are presented in Section 4. Section 5 describes the error analysis corresponding to our estimations. The discussion related to our study are presented in Section 6. We summarise the conclusions in Section 7.
\begin{figure*} 
\begin{center} 
\includegraphics[height=6.in,width=6.0in]{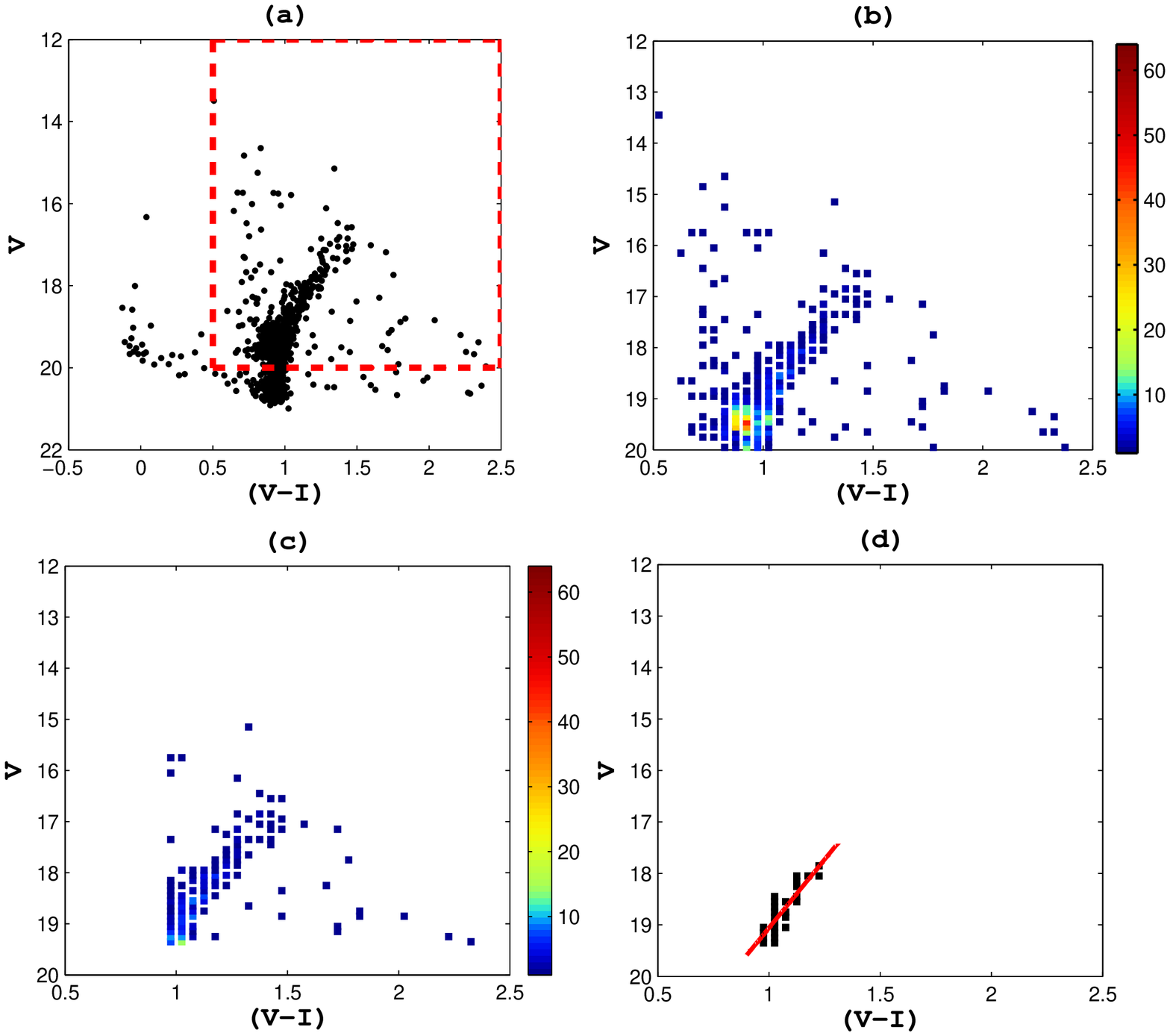}
\vskip 6cm
\caption{\small (a) The V versus (V$-$I) CMD of an OGLE III region at (8.39$^\circ$, $-$72.51$^\circ$), of size (8.88$\times$8.88) sq. arcmin, with N=952 stars (black filled circles). The stars within the rectangle (red dashed line) belongs to the evolved part of the CMD. (b) Density diagram of the evolved part of CMD, where the CMD bins are colour coded based on the number of stars contained in them, as denoted in the colour bar. (c) Density diagram after giving a colour-magnitude cut at the peak value of RC distribution. (d) Density diagram showing CMD bins that have $\ge$ 3 stars (black) . Straight line fit to these bins representing the RGB, after 3$\sigma$ clipping, is shown as a red solid line. The estimated parameters are: $|$slope$|$= 5.31$\pm$0.66, $r$=0.85, and $N_p$= 26.}  
\label{fig:fig01}
\end{center} 
\end{figure*}
\begin{table*}
{\small
\caption{Sub-division of OGLE III regions:}
\label{table:tab1}
\begin{tabular}{|c|c|c|c|c|c|c|c|}
\hline \hline
Sl. no. & No. of  & No. of  & No. of     & No. of     & No. of             & Area of        &  Number of       \\
        & stars   & regions & division   & division   & sub-divisions      & a sub-division &  subregions      \\
        &         & (a)     & along RA   & along Dec  & (d=b$\times$c)     & (arcmin sq.)   &  (a$\times$d)    \\
        &         &         & (b)        & (c)        &                    &                &                  \\
\hline\hline
1  & 0 $<$ N $\le$ 1600      & 266 & 1 & 1 & 1  & (8.88$\times$8.88) & 266 (black)     \\
2  & 1600 $<$ N $\le$ 3600   & 160 & 2 & 1 & 2  & (4.44$\times$8.88) & 320 (brown)    \\
3  & 3600 $<$ N $\le$ 5000   & 51  & 3 & 1 & 3  & (2.96$\times$8.88) & 153 (red)       \\
4  & 5000 $<$ N $\le$ 7200   & 58  & 2 & 2 & 4  & (4.44$\times$4.44) & 232 (orange)    \\
5  & 7200 $<$ N $\le$ 11000  & 39  & 3 & 2 & 6  & (2.96$\times$4.44) & 234 (yellow)    \\
6  & N $>$ 11000           & 15  & 4 & 2 & 8  & (2.22$\times$4.44) & 120 (dark green)\\
\hline     
\end{tabular}
\vskip 1.0ex
\begin{minipage} {180mm}
{Note: The table describes the 6 binning criteria used to sub-divide OGLE III regions. For each criteria, the second column denotes the limit on total number of stars (N) within a region. The third column gives the number of regions, having N, within that specified limit. Columns four and five specify the number by which a region is binned along RA and Dec respectively. Column six thus gives the total number of subregions, a single region is binned into. Whereas, the seventh column gives the area of each such subregion. The last (eighth) column denotes the total number of subregions corresponding to each of the 6 sub-division criteria. The colours adjacent to the numbers are used to denote them in Figure \ref{fig:fig04} and \ref{fig:fig05}.}
\end{minipage}
}
\end{table*} 


\begin{figure*}
\centering
\begin{minipage}[b]{0.45\linewidth}
\includegraphics[height=3.0in,width=3.0in]{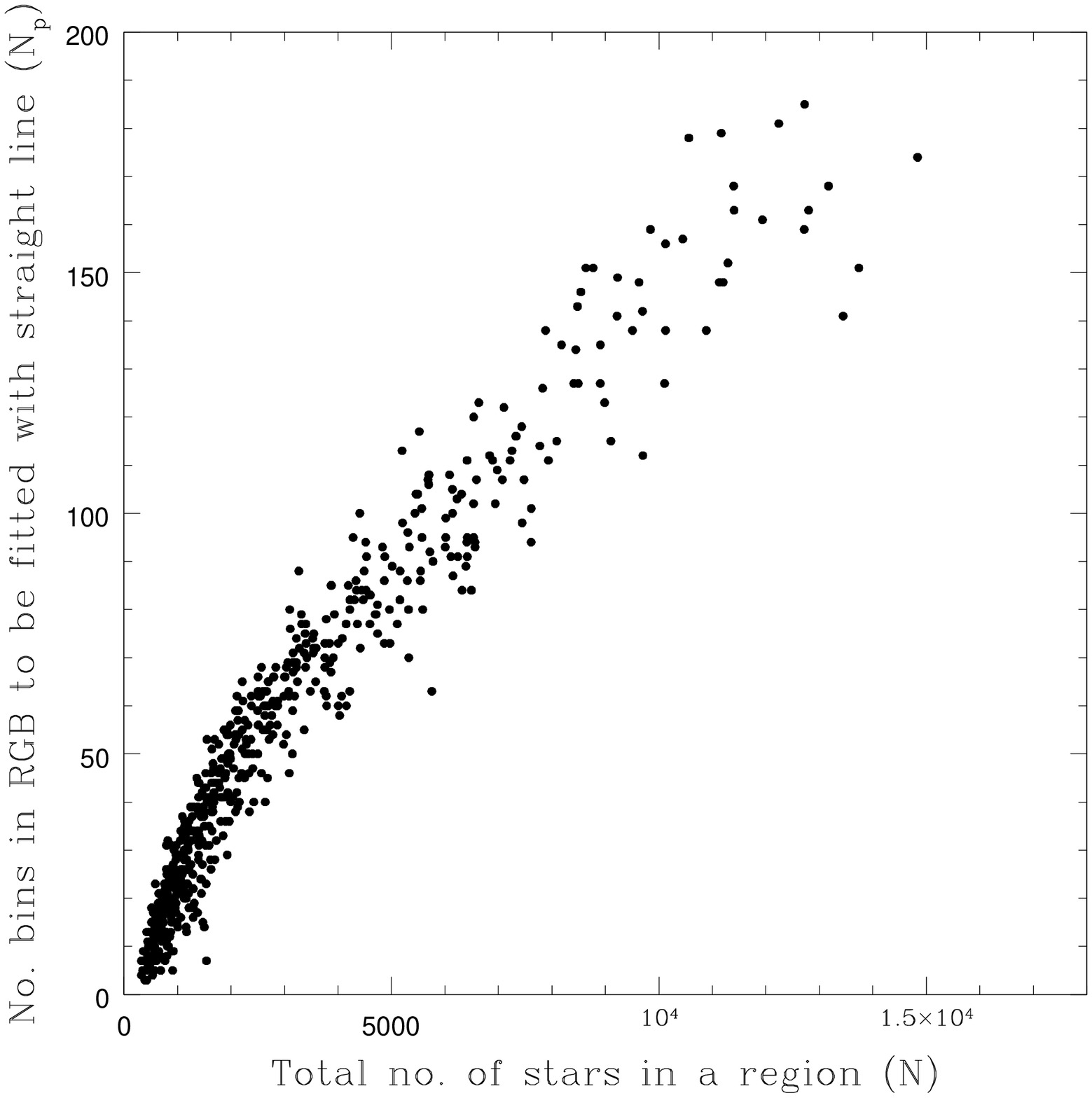}
\vskip 2cm
\caption{\small Plot of number of bins in RGB to be fitted with straight line ($N_p$) versus the total number of stars ($N$) for OGLE III subregions, after initial area binning}
\label{fig:fig02}
\end{minipage}
\quad
\begin{minipage}[b]{0.45\linewidth}
\includegraphics[height=3.0in,width=3.0in]{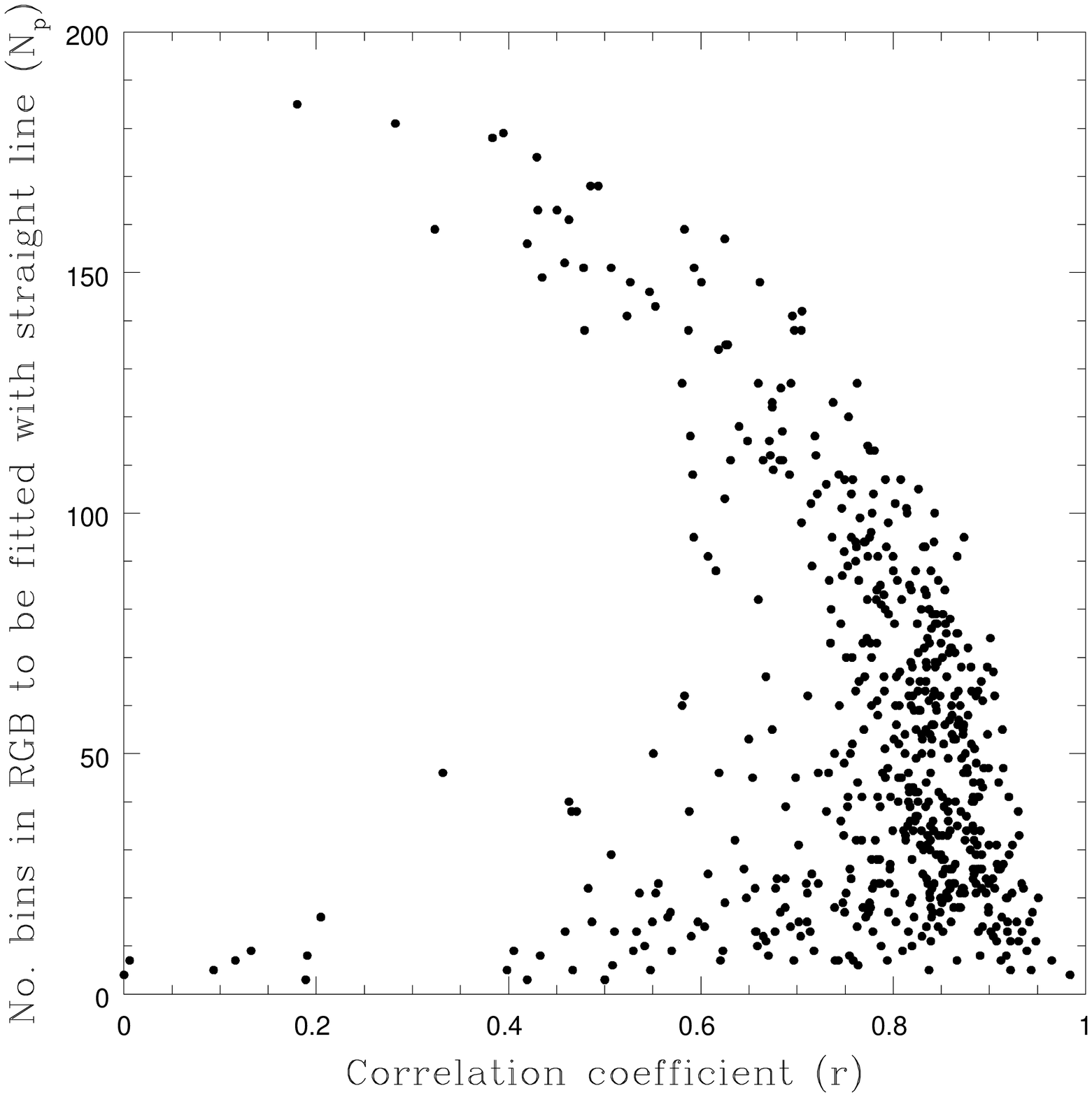}
\vskip 2cm
\caption{\small Plot of  number of bins in RGB to be fitted with straight line ($N_p$) versus correlation coefficient ($r$) for OGLE III subregions, after initial area binning}
\label{fig:fig03}
\end{minipage}
\end{figure*} 


\begin{figure*}
\centering
\begin{minipage}[b]{0.45\linewidth}
\includegraphics[height=3.0in,width=3.0in]{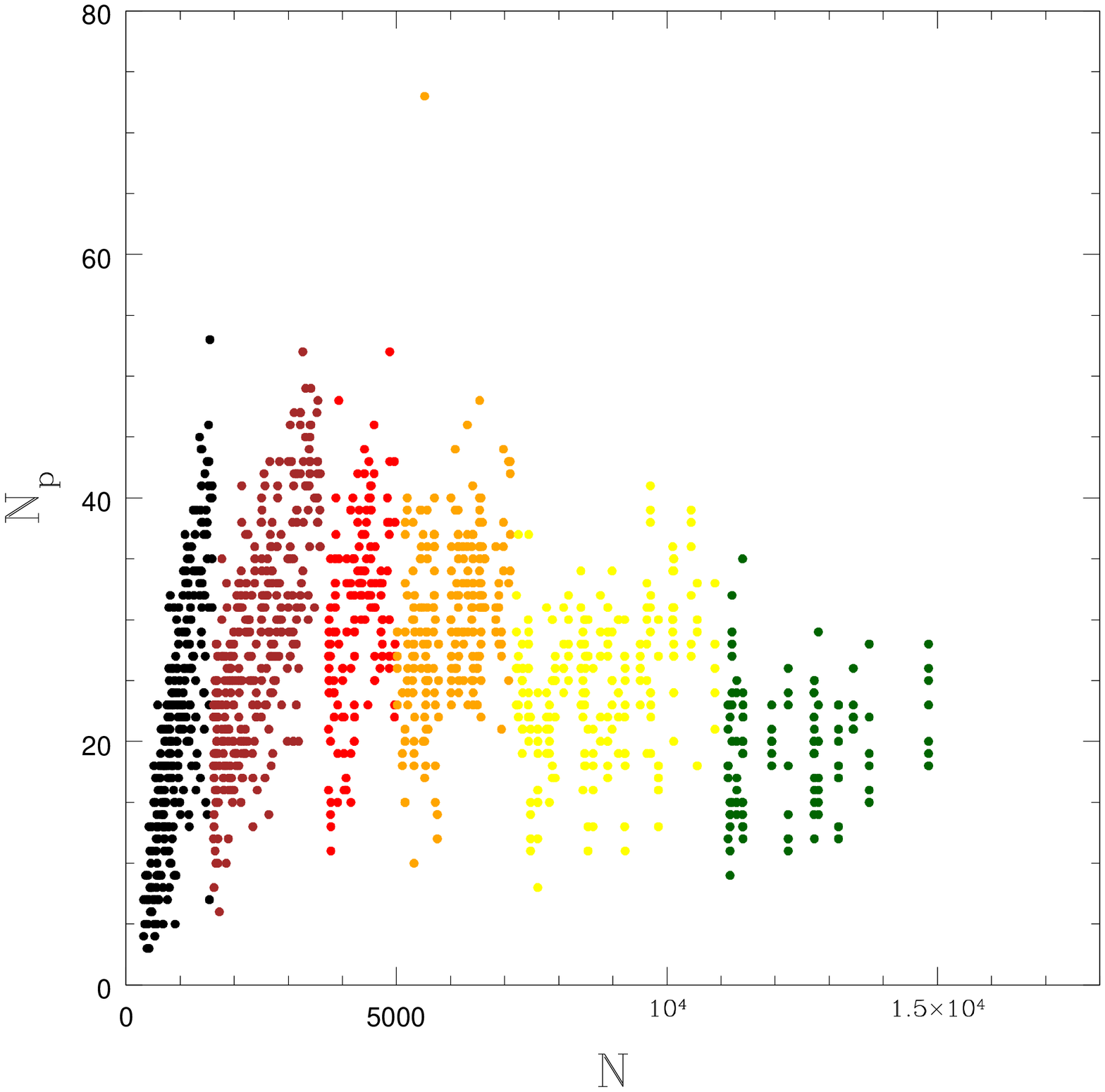}
\vskip 2cm
\caption{\small Plot of $N_p$ versus $N$ for OGLE III subregions, after finer area binning. The colours correspond to the six different bin areas, as mentioned in the eighth column of Table \ref{table:tab1}.}
\label{fig:fig04}
\end{minipage}
\quad
\begin{minipage}[b]{0.45\linewidth}
\includegraphics[height=3.0in,width=3.0in]{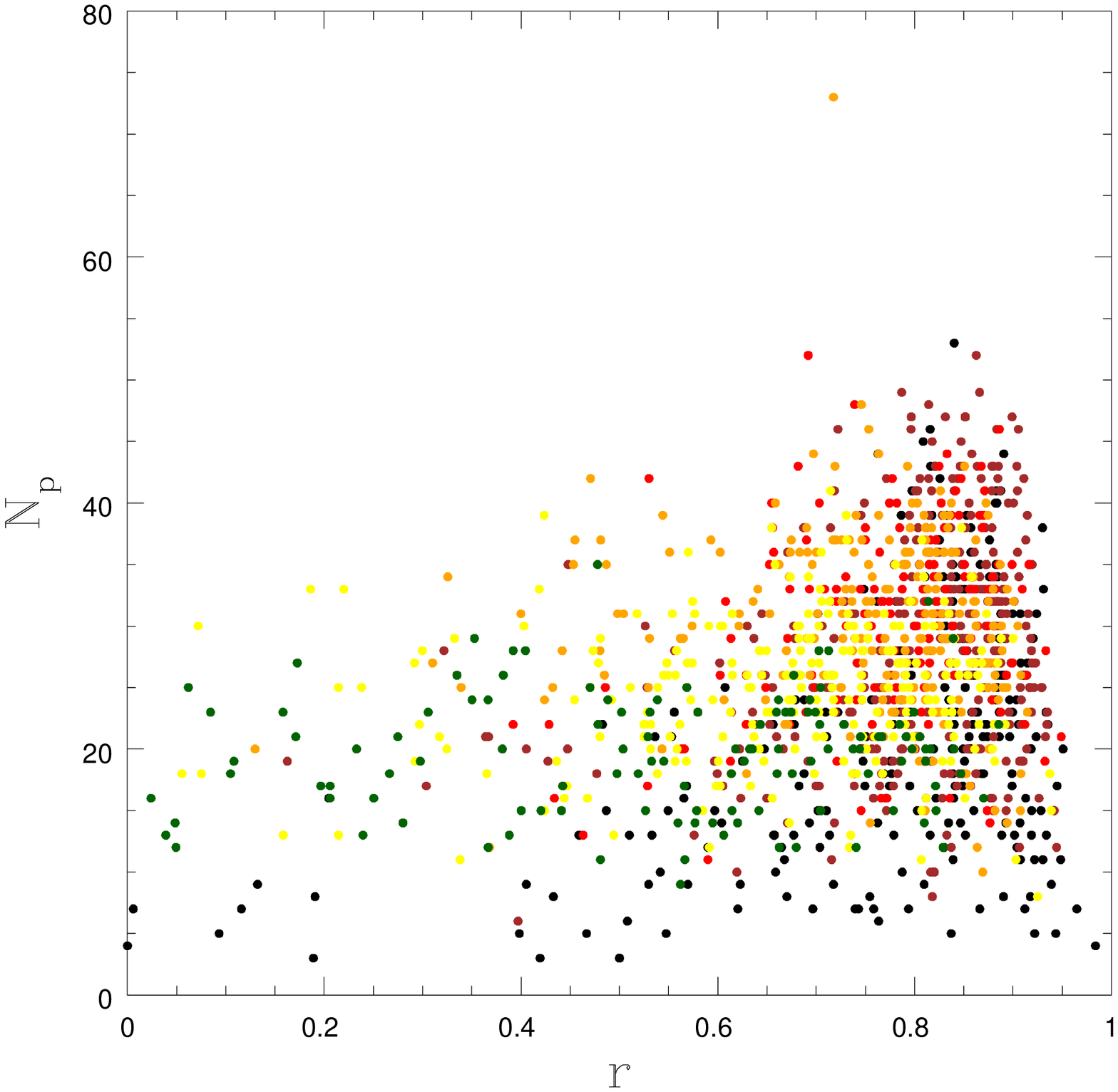}
\vskip 2cm
\caption{\small Plot of $N_p$ versus $r$ for OGLE III subregions, after finer area binning. The colours correspond to the six different bin areas, as mentioned in the eighth column of Table \ref{table:tab1}.}
\label{fig:fig05}
\end{minipage}
\end{figure*}
\section{Data}
For this study we have used two large photometric surveys of the SMC, OGLE III and the MCPS. The OGLE III survey covered a total area of about 14 square degrees \citep{Udalski+2008AcAOGLEIIISMC} and presented calibrated V and I band photometry of about 6.2 million stars. OGLE III covers the central region as well as the eastern and western regions of the inner SMC till a radius of $\sim$ 2.5$^{\circ}$ from the SMC centre. The MCPS provided the catalogue for over 5 million stars in the SMC, within the central 18 square degrees of the SMC \citep{Zaritsky+2002AJMCPSSMC}. The survey presented the photometric as well as extinction maps in U, B, V, and I passbands for the SMC. The MCPS survey is of relatively lower resolution (0$^{''}$.70 pixel$^{-1}$) compared to the OGLE III (0$^{''}$.26 pixel$^{-1}$) survey. We only consider stars with photometric error less than 0.15 mag in the V and I passbands in both the data sets. The MCPS has more coverage of the northern and southern regions of the central SMC, whereas the OGLE III survey has more coverage of the eastern and western regions. The two surveys thus complement each other in terms of area covered.

\section{OGLE III Analysis}

We use the the slope of the RGB in the CMD of a small region in the galaxy as an indicator of the mean metallicity of the region. A region in a galaxy with metal-rich stars is expected to have a shallower RGB slope when compared to a relatively metal-poor region. The dependence of slope of the RGB on metallicity is well known and has been studied primarily for homogeneous populations, i.e., star clusters\citep{DaCosta&Armandroff1990AJstandard, Kuchinski+1995AJ-IRarray}. In Paper I we applied this concept for the first time to field stars, which are heterogeneous with respect to age and metallicity. The RGB slope of a field region will correspond to the metallicity of the dominant RGB population of the region. Section 3 of Paper I describes in detail the salient features of the robust process developed to identify and estimate the RGB slope consistently within the CMDs of small regions within the galaxy, independent of differential reddening. 

The OGLE III observed region is binned into 656  small regions, each of dimension (8.88 $\times$ 8.88) sq. arcmin in RA and Dec. We identify the RGB by (i) segregating it from other evolutionary stages, (ii) using the location of Red Clump (RC) stars to locate the base of the RGB, and (iii) estimating the slope of the RGB in CMDs of all these small regions. According to \cite{Smitha&Purni2009} most of the SMC regions have a large number of RC stars and their distribution can be well-identified in the CMDs of the regions. The use of the peak of the RC as RGB base allows us to uniquely and consistently define a location in the RGB and identify it uniformly in CMDs of all the regions. Thus, even if this is not the actual base of the RGB, the part of RGB used for slope estimation is made uniform for all location.

Similar to Paper I, the RGB slope estimation procedure was tested and verified for a large number of SMC subregions before validating. We briefly mention the primary steps involved in the process (for detailed steps we direct the readers to Section 3 of Paper I):

(1) Excluding the Main Sequence (MS) and including mainly the evolved portion of the CMD with stars having 0.5 $<$ (V$-$I) $\le$ 2.5 mag and 12.0 $\le$ V $<$ 20.0 mag (Figure \ref{fig:fig01}a). (2) Constructing a density diagram to identify the location of RC stars and assume the peak (V$-$I) and V of the RC distribution as the base of the RGB (Figure \ref{fig:fig01}b). This location changes with the reddening, along with the location of the RGB, and is used to identify the RGB similarly in all the regions. (3) Removing bluer and fainter bins with respect to the RC peak by giving a cut in colour and magnitude corresponding to the RC peak (Figure \ref{fig:fig01}c). The extracted part of the CMD is dominated by RGB stars, but contaminated with some other evolutionary phases (AGBs, etc.) to a lesser extent. (4) Finally, in order to identify the RGB unambiguously, and to reduce the scatter in the RGB and eliminate the other evolutionary phases, we consider only those colour-magnitude bins that contain a minimum number of 3 stars. The chosen criterion eliminates the brighter part of the RGB, typically sampling the RGB from the RC peak up to 2 magnitude brighter in most of the regions. As shown in Figure \ref{fig:fig01}(d), the selected part of the RGB thus appears to be more or less a straight line, without the curved brighter part. These bins, thus representing the RGB, are fitted with a straight line and the slope is estimated using the method of least square fit (3-$\sigma$ clipping with a single iteration).

The total number of stars in a particular region is given by N. We define $N_p$, as the number of  CMD bins (with number of stars in each bin $\ge$ 3) representing the RGB. The slope of RGB is negative, but we denote it by its absolute value as $|$slope$|$, and the error in slope as $\sigma_{slope}$. We also express the correlation coefficient by its absolute value, as $r$. Thus, we have calculated $N_p$, $|$slope$|$, $\sigma_{slope}$, and $r$ for each of the regions. 


\begin{figure*}
\centering
\begin{minipage}[b]{0.45\linewidth}
\includegraphics[height=3.0in,width=3.0in]{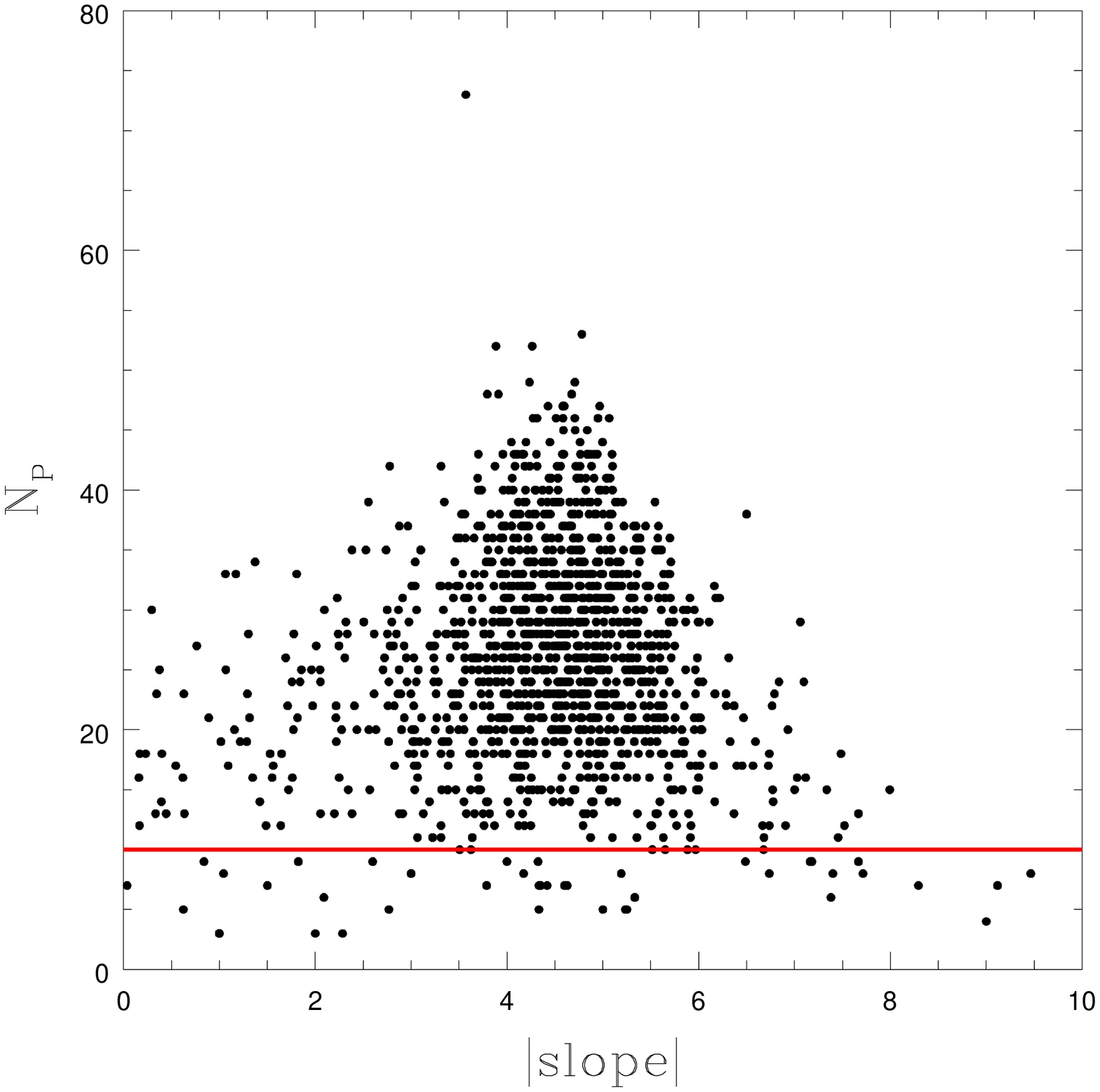}
\vskip 2cm
\caption{\small Plot of $N_p$ versus $|$slope$|$ for OGLE III subregions. The red line at $N_p$ = 10 denotes the cut-off decided to exclude regions with poorly populated RGB.}
\label{fig:fig06}
\end{minipage}
\quad
\begin{minipage}[b]{0.45\linewidth}
\includegraphics[height=3.0in,width=3.0in]{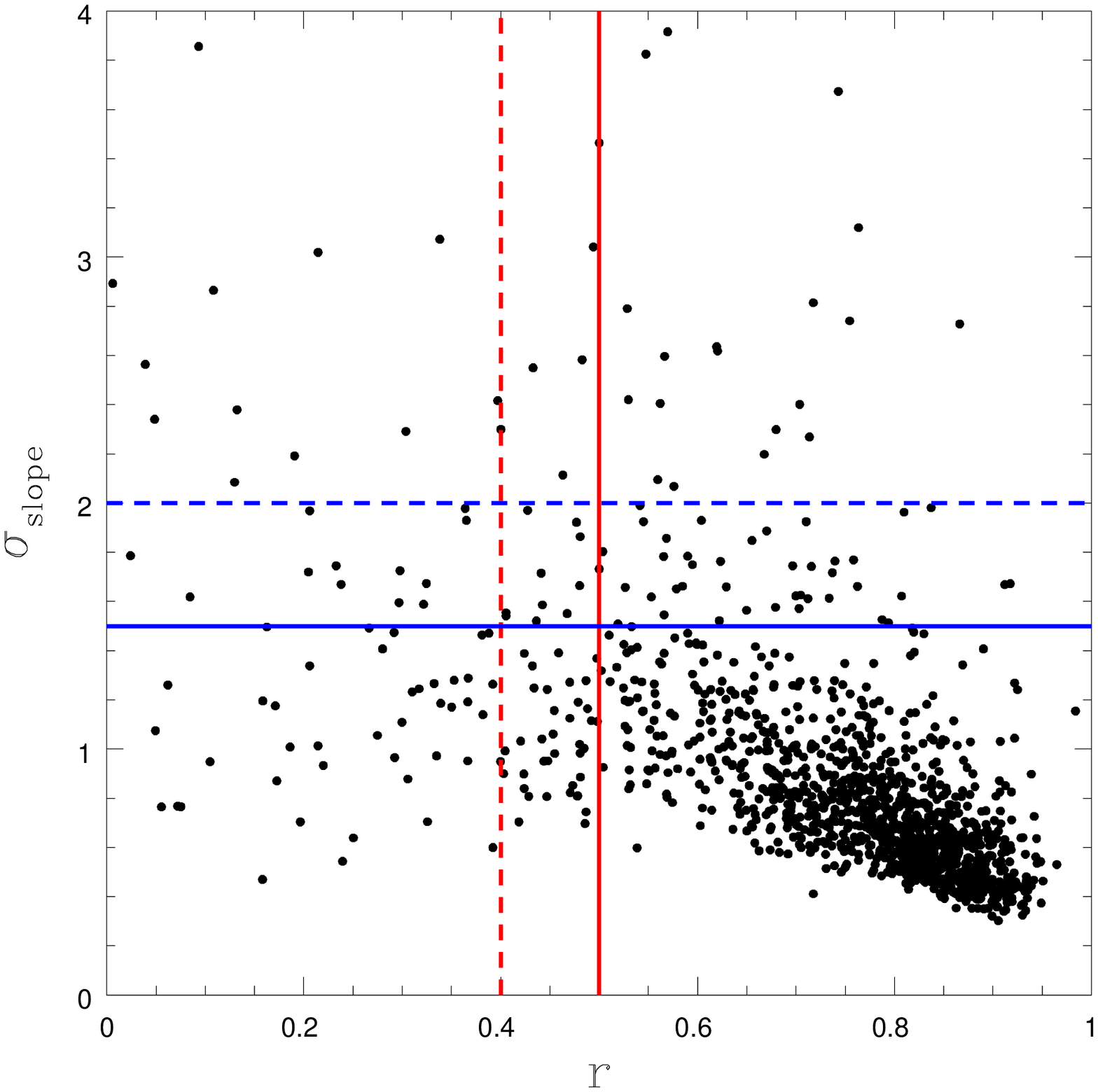}
\vskip 2cm
\caption{\small Plot of $\sigma_{slope}$ versus $r$ for OGLE III subregions. The blue dashed and solid lines corresponds to the cut-off criteria on $\sigma_{slope}$ at 2.0 and 1.5 respectively. The red dashed and solid lines denote the cut-off corresponding to $r$ at 0.4 and 0.5 respectively.}
\label{fig:fig07}
\end{minipage}
\end{figure*} 


\begin{figure*} 
\begin{center} 
\includegraphics[height=4.0in,width=4.0in]{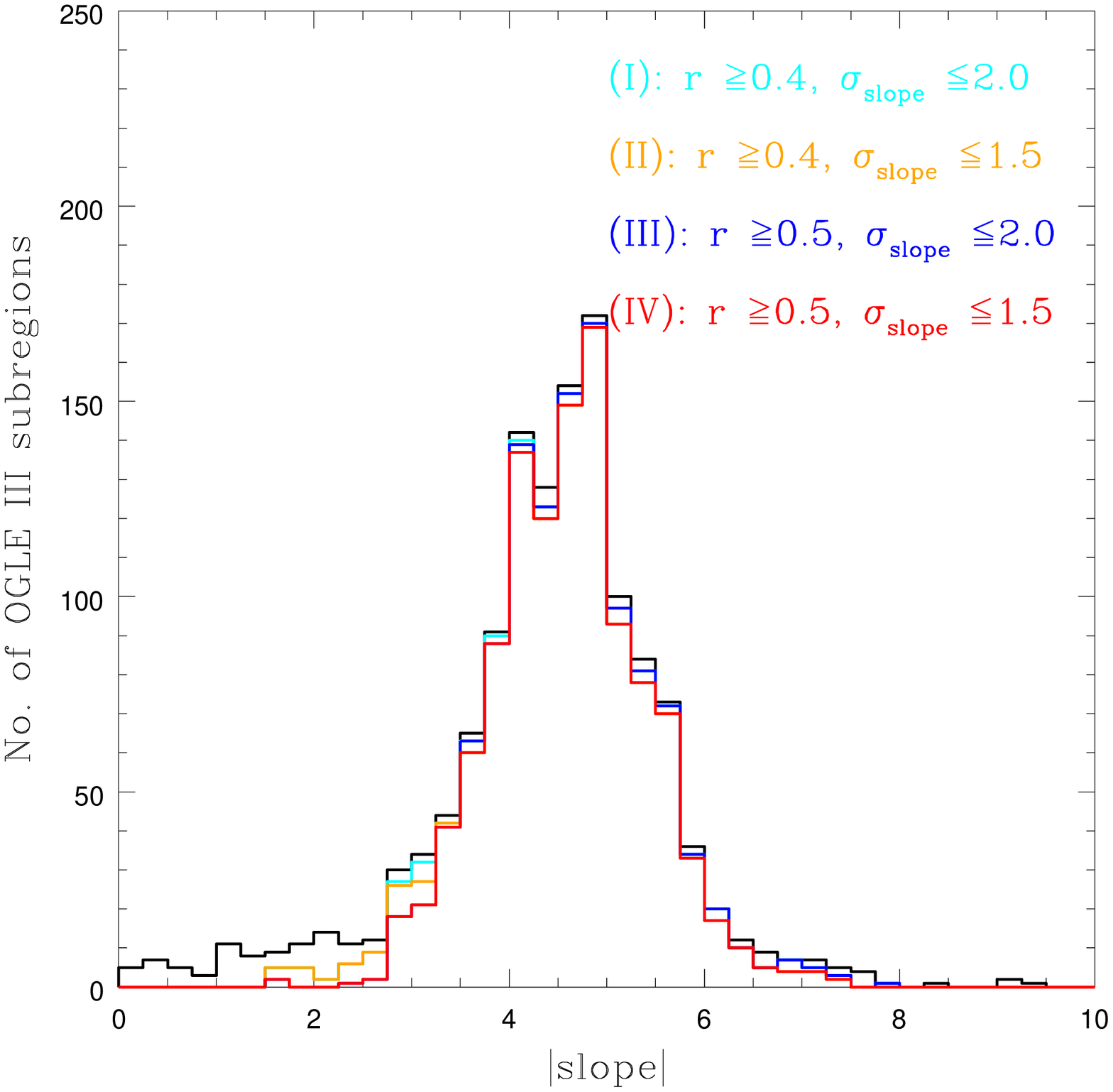}
\vskip 2cm
  \caption{Histogram of $|$slope$|$ for OGLE III subregions estimated for all the four cut-off criteria ($(I)$ in cyan, $(II)$ in orange, $(III)$ in blue, and $(IV)$ in red). $N_p$ $\ge$ 10 for all these four cases. The black solid line shows the distribution of $|$slope$|$ with no cut-off corresponding to all OGLE III subregions.}
\label{fig:fig08}  
\end{center} 
\end{figure*}


The above method is found to work consistently for most of the regions, except when there is a large variation in reddening and/or multiple dominant stellar population. The central regions which have higher stellar density show a scatter in the RC and RGB thus providing poor estimation of slope (lower $r$). In order to probe the effect on the fit due to over-populated  CMDs, Figure \ref{fig:fig02} shows a plot of $N_p$ versus N. The figure shows that as N increases, $N_p$ also increases, suggesting that, in general, the denser areas have well-populated RGBs. In Figure \ref{fig:fig03}, the correlation between $N_p$ and $r$ is shown, where it is seen that the regions with high $N_p$ (i.e. correspondingly higher N) have lower $r$ ($<$ 0.50), suggesting a poor fit/estimation of slope. This is similar to Figure 2 and 3 of Paper I, where we note that adopting similar areas for all regions across the galaxy leads to poor estimation of slopes for regions which have high stellar density. Inspection of such regions suggests that the broad RGB in such regions are likely caused by small-scale variation in reddening and/or multiple dominant population. 

We further showed in Paper I that finer subdivision of regions based on stellar density can help get rid of this problem. To achieve this, we adopted 6 binning criteria for the OGLE III regions, based solely on star counts as shown in Table \ref{table:tab1}. It also lists the total number of sub-regions extracted under each criterion, along with their corresponding areas. The area of the largest sub-regions is (8.88 $\times$ 8.88) sq. arcmin and the smallest is (2.22$\times$4.44) sq. arcmin. There are a total of 1322 sub-regions analysed after finer area binning. Figure \ref{fig:fig04} and \ref{fig:fig05} show the plot of $N_p$ versus N, and $N_p$ versus $r$ after finer area binning. There are now more regions with $r$ $>$ 0.5, and the value of $N_p$ is confined to lower and similar values for all the 6 binning criteria. We still have some regions with low value of $r$, which we found are due to either poorly-defined or very broad RGB. As found by \cite{Smitha&Purni2009,Smitha&Purni2012} the line of sight (LOS) depth of the SMC is almost constant within the inner SMC (our studied region). Thus, our method of RGB slope estimation is expected not to be strongly affected by LOS depth. However, there are regions with variation in depth in the central and north-east of the SMC. This variation in depth could contribute to poor RGB slope estimation in a minority of sub-regions.

To carry forward our analysis by eliminating regions with poor fit, we choose cut-off criteria on the estimated parameters $N_p$, $r$ and $\sigma_{slope}$ (Paper I). Figures \ref{fig:fig06} and \ref{fig:fig07} show the plot of $N_p$ versus $|slope|$ and $\sigma_{slope}$ versus $r$ respectively. In Figure \ref{fig:fig06} regions with $N_p$ $<$10 and very large value for $|$slope$|$ signify a sparsely-populated RGB. Hence, the large slope value may be an artefact. Thus, to exclude regions with poorly populated RGB we select only those regions that have $N_p$ $\ge$ 10, implying that the fitted RGB should at least have 30 stars. Figure \ref{fig:fig07} shows that most of the regions have $r$ in the range 0.4-0.95, and $\sigma_{slope}$ in the range 0.5-2.0. By contrast, there is a large scatter observed for regions with $\sigma_{slope}$ $>$ 2.0 and $r$ $<$ 0.4. Also, the clumpiest part of the figure is seen for $\sigma_{slope}$ $<$ 1.5 and $r$ $>$ 0.5. Thus, based on Figure \ref{fig:fig07} we consider four different cut-off criteria in terms of $r$ and $\sigma_{slope}$. These are (with $N_p$ $\ge$ 10 for all criteria) as follows: 
\begin{itemize}
\item criteria (I): $r$ $\ge$ 0.4 and $\sigma_{slope}$ $\le$ 2.0.
\item criteria (II): $r$ $\ge$ 0.4 and $\sigma_{slope}$ $\le$ 1.5.
\item criteria (III): $r$ $\ge$ 0.5 and $\sigma_{slope}$ $\le$ 2.0.
\item criteria (IV): $r$ $\ge$ 0.5 and $\sigma_{slope}$ $\le$ 1.5.
\end{itemize}
We have plotted a histogram of the slope for all the four cut-off criteria along with the original distribution with no cut-off in Figure \ref{fig:fig08}. Comparing this with the RGB slope distribution of the LMC for OGLE III data (Figure 8 of Paper I), we find that the distributions appear to be different from each other in peaks and width, apart from the fact that SMC has fewer subregions compared to the LMC. The SMC slope distribution shows two prominent peaks, a primary at $\sim$ 5 and a secondary at $\sim$ 4, whereas the LMC distribution is peaked primarily at $\sim$ 3.5, indicating a metal-rich case compared to the SMC. Figure \ref{fig:fig08} shows that as the cut-off criteria becomes stringent from (I) to (IV), the regions with lower values of slopes ($<$3) are mostly removed. However, the effect of increasing the cut-off for $r$ for similar values of $\sigma_{slope}$ is more than that of lowering the cut-off for $\sigma_{slope}$ for same value of $r$. This effect is similar to that we noticed for the LMC, where slope with values $<$2 were mostly removed. Thus, our last criteria (IV) is the most stringent one and the first one is the most relaxed one for selecting regions with best fits. The range of RGB slope for the SMC is found to be primarily within $\sim$ 3--6 (with a few values lower than 3 and greater than 6). The LMC on the other hand had a different slope range, primarily varying between $\sim$ 2--6.

\begin{table*}
{\small
\caption{Calibrators for OGLE III slope-metallicity relation:}
\label{table:tab2}
\begin{tabular}{|c|c|c|c|c|c|c|c|c}
\hline \hline
RA$^{\circ}$ & Dec$^{\circ}$ & $|$slope$|$ & $\sigma_{slope}$ & $r$ & Mean [Fe/H] & Standard error \\
             &               &             &                  &     &    (dex)    &  of mean [Fe/H]\\
             &               &             &                  &     &             &                \\
\hline \hline
         11.82 & -71.88 &     4.93 & 0.32 & 0.93 & -0.95  & 0.07 & \\
         13.35 & -74.55 &     5.00 & 0.50 & 0.86 & -1.13  & 0.06 & \\
         14.57 & -71.49 &     4.53 & 0.39 & 0.88 & -0.88  & 0.06 & \\
         14.08 & -71.50 &     5.11 & 0.81 & 0.75 & -1.16  & 0.07 & \\
         18.88 & -72.38 &     4.91 & 0.57 & 0.86 & -0.95  & 0.07 & \\
         18.39 & -72.23 &     5.40 & 0.61 & 0.84 & -1.02  & 0.05 & \\
         18.86 & -72.23 &     4.98 & 0.55 & 0.88 & -1.07  & 0.07 & \\
         18.87 & -72.07 &     6.37 & 0.67 & 0.90 & -1.08  & 0.07 & \\
          8.90 & -72.50 &     4.97 & 0.50 & 0.87 & -1.00  & 0.06 & \\
          8.13 & -74.15 &     4.84 & 0.54 & 0.81 & -1.13  & 0.07 & \\
         13.13 & -71.88 &     4.74 & 0.47 & 0.89 & -0.99  & 0.07 & \\
         11.92 & -72.19 &     4.73 & 0.39 & 0.89 & -0.87  & 0.07 & \\
         13.53 & -71.93 &     4.71 & 0.38 & 0.91 & -0.88  & 0.07 & \\
         13.77 & -71.93 &     4.96 & 0.49 & 0.85 & -0.95  & 0.07 & \\
         14.91 & -71.64 &     4.83 & 0.79 & 0.84 & -0.93  & 0.06 & \\
         13.72 & -71.80 &     3.78 & 0.78 & 0.71 & -0.92  & 0.07 & \\
         13.96 & -71.80 &     4.11 & 0.56 & 0.80 & -0.90  & 0.07 & \\
         14.19 & -71.80 &     4.96 & 0.70 & 0.80 & -0.83  & 0.07 & \\
         17.10 & -72.08 &     4.63 & 0.63 & 0.87 & -0.95  & 0.07 & \\
         17.47 & -72.52 &     5.20 & 0.69 & 0.81 & -1.12  & 0.07 & \\
         11.35 & -74.28 &     4.99 & 0.52 & 0.89 & -0.99  & 0.07 & \\
         16.66 & -72.82 &     4.51 & 0.59 & 0.84 & -0.80  & 0.06 & \\
         16.66 & -72.67 &     4.52 & 0.54 & 0.86 & -0.83  & 0.07 & \\
         16.82 & -72.67 &     3.55 & 0.51 & 0.80 & -0.84  & 0.06 & \\
         16.99 & -72.67 &     5.10 & 0.81 & 0.81 & -1.00  & 0.07 & \\
         16.66 & -72.52 &     3.76 & 0.67 & 0.71 & -0.93  & 0.07 & \\
         13.14 & -72.15 &     5.31 & 0.60 & 0.88 & -0.90  & 0.07 & \\
         16.27 & -72.71 &     4.26 & 0.65 & 0.79 & -0.93  & 0.07 & \\
         10.13 & -73.14 &     4.55 & 0.54 & 0.81 & -1.09  & 0.07 & \\
         10.38 & -73.14 &     4.76 & 0.64 & 0.80 & -0.90  & 0.07 & \\
         10.46 & -72.92 &     4.12 & 0.57 & 0.76 & -0.77  & 0.07 & \\
         12.79 & -73.54 &     4.01 & 0.62 & 0.79 & -0.94  & 0.06 & \\
         11.35 & -72.77 &     4.60 & 0.95 & 0.75 & -0.82  & 0.07 & \\

\hline     
\end{tabular}
\begin{minipage} {180mm}
\vskip 1.0ex
{Note: The table lists out the 33 calibrators used to construct the slope-metallicity relation for OGLE III data. The central (RA, Dec) corresponding to each calibrator is listed down in first and second column respectively. The third, and fourth column gives the estimated slope and its associated error respectively, whereas, the correlation coefficient for each case is specified in the fifth column. The sixth column denotes the mean [Fe/H] estimated for each calibrator using spectroscopic results of \cite{Dobbie+2014MNRAS-papII}. The standard error of mean [Fe/H] is  mentioned in the last column. It is calculated as 0.15/$\sqrt{n}$, where n is the number of spectroscopically studied RGs located within the area of the subregion.}
\end{minipage}
}
\end{table*} 

\begin{figure*} 
\begin{center} 
\includegraphics[height=4in,width=4in]{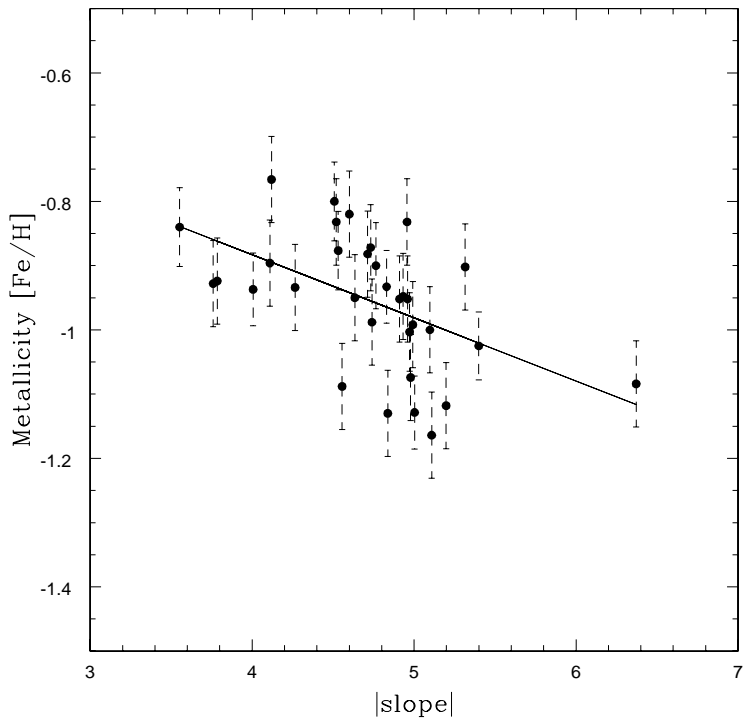}
\vskip 2cm
\caption{Plot of [Fe/H] versus $|$slope$|$ for OGLE III data. The points denote our subregions whose mean [Fe/H] has been found using RGs from \citealt{Dobbie+2014MNRAS-papII}, with the solid line denoting a linear relation between them. The error bar (dashed line) shown for each point is the standard error of mean [Fe/H].}
\label{fig:fig09}
\end{center} 
\end{figure*}

\subsection{Calibration of RGB slope to Metallicity}

After estimating the RGB slope of sub-regions within the SMC, the next task is to convert the slope to metallicity. As the slope of the RGB is a measure of the average metallicity of the region, we use the spectroscopically estimated average metallicity of red giants in that region to build the required relation. We define this relation based on the spectroscopically determined mean metallicity for several regions.

We used metallicities of 3037 field red giants from \cite{Dobbie+2014MNRAS-papII} to calibrate the OGLE III RGB slope map. The reasons behind using their study are: (1) They derived metallicities for the same population as our study, the RGBs; (2) Our RGB slope has a range of values and hence requires a corresponding range in metallicity to formulate a relation between the two. To our knowledge, Dobbie et al.'s work is the most extensive spectroscopic study of SMC field RGs in terms of the number of stars and the spatial area covered. Also, their study spans a large range ($\sim$ $-$0.5 dex to about $-$2.5 dex) in metallicity. We calculate the mean metallicity for a sub-region by averaging over the Dobbie et al.\ metallicities within its area. While doing so, we consider stars lying within twice the standard deviation about the mean metallicity. To ensure good calibration we consider only those sub-regions that have $r$ higher than 0.70, $\sigma_{slope}$ lower than 1.0, and contain spectroscopic metallicity estimate of at least 5 RGs. These cut-off conditions used are similar to that used for calibrating the LMC metallicity map (Section 3.1 of Paper I).

The sub-regions used for calibration are plotted in the metallicity versus $|$slope$|$ plane. The trend shows that with increasing value of $|$slope$|$, the metallicity decreases. We estimate a linear relation between them by fitting a straight line using least square fit with 2$\sigma$ clipping. In Figure \ref{fig:fig09} the final 33 sub-regions used for the slope-metallicity calibration are shown, and these values are also tabulated in Table \ref{table:tab2}. The slope-metallicity relation estimated is given by:
\begin{equation} \label{eq:1}
[Fe/H]=(-0.098\pm 0.029)\times|slope|+ (-0.490\pm 0.140);
\end{equation}
with $r$=0.51. The estimated slope ($-$0.098) of this RGB slope-metallicity relation is shallower when compared to that of the LMC in Paper I ($-$0.137, their Equation 1). The y-intercept ($-$0.489) of the slope-metallicity relation is metal poor for the SMC as compared to the LMC (0.092) owing to the galaxy being relatively metal poor. It is to be noted that for the SMC we have been able to establish the relation with a larger number of data points (33) than for the LMC (16). Also, the range of $|$slope$|$ for these points ($\sim$ 3.2 $\la$ $|$slope$|$ $\la$ 6.2) almost covers the entire slope distribution of the SMC (Figure \ref{fig:fig08}). Whereas, for the LMC the slope range of the calibrators (3.2 $\la$ $|$slope$|$ $\la$ 4.6) covered mainly the peak of the RGB slope distribution (Figure 8 of Paper I). Thus, we have successfully established a slope-metallicity relation for the SMC for the OGLE III data set.


\begin{figure*} 
\begin{center} 
\includegraphics[height=5in,width=7in]{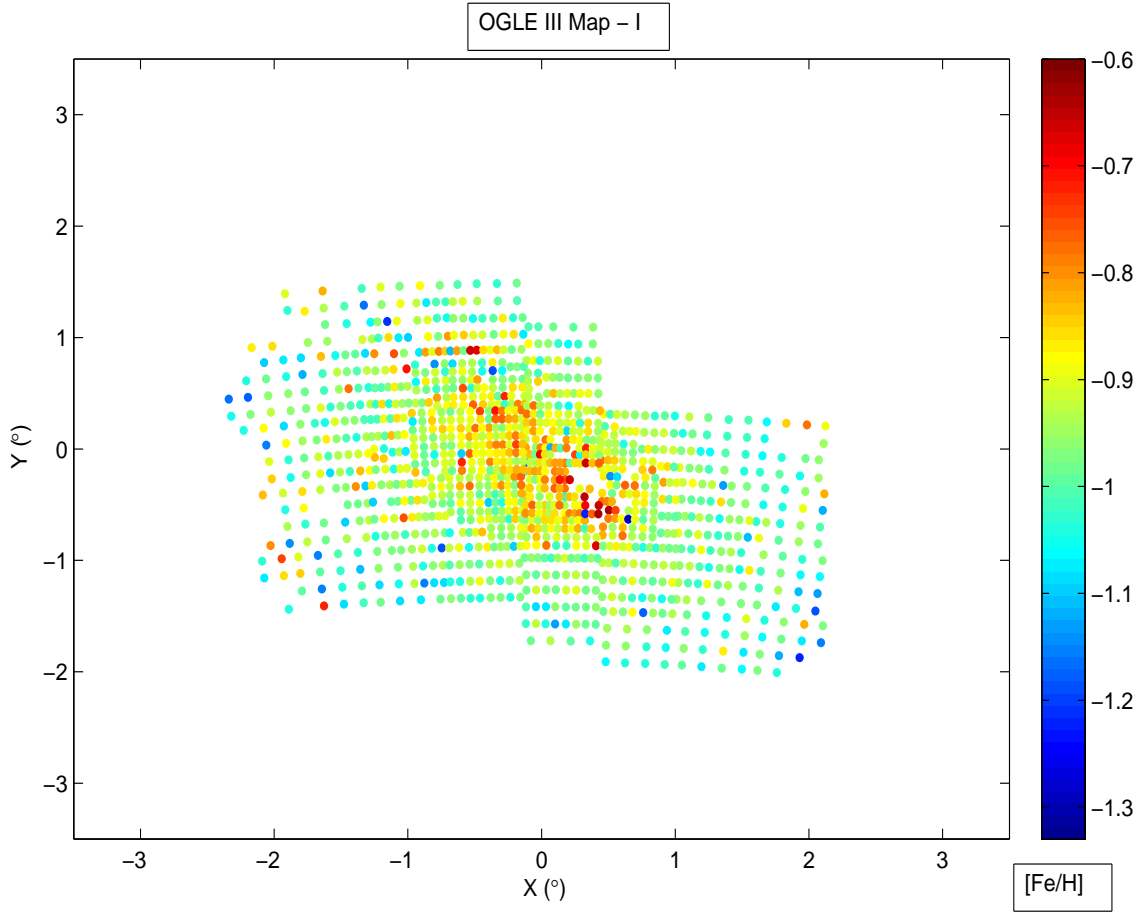}
\vskip 6cm
\caption{OGLE III metallicity map with cut-off criteria $(I)$: $N_p$ $\ge$ 10, $r$ $\ge$ 0.4 and $\sigma_{slope}$ $\le$ 2.0.}
\label{fig:fig10} 
\end{center} 
\end{figure*}

\begin{figure*} 
\begin{center} 
\includegraphics[height=5in,width=7in]{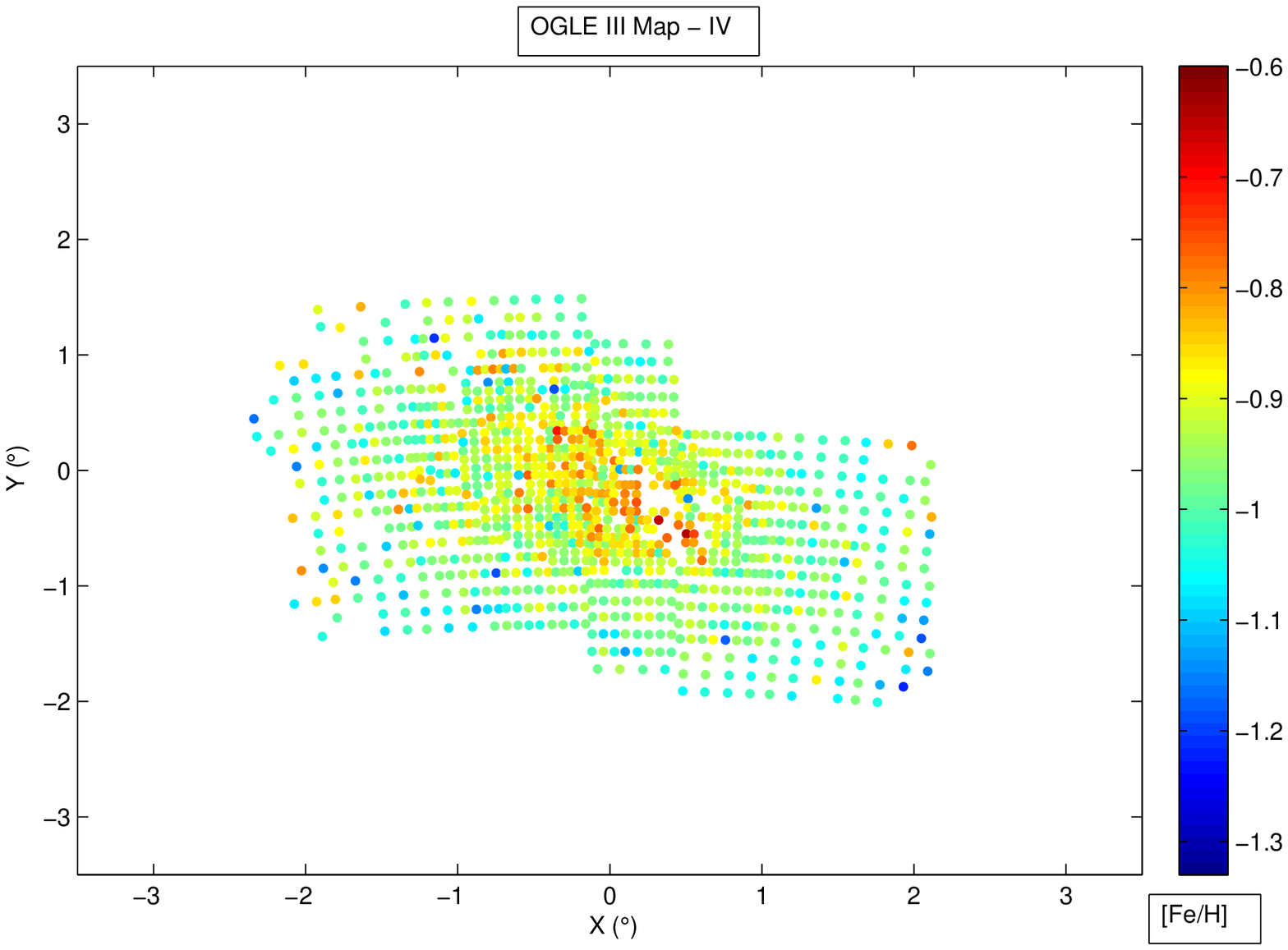}
\vskip 6cm
\caption{OGLE III metallicity map with cut-off criteria $(IV)$: $N_p$ $\ge$ 10, $r$ $\ge$ 0.5 and $\sigma_{slope}$ $\le$ 1.5.}
\label{fig:fig11} 
\end{center} 
\end{figure*}
\subsection{OGLE III Metallicity Map}

We use Equation \ref{eq:1} to convert all the RGB slopes to metallicity. The OGLE III metallicity maps are shown in Figure \ref{fig:fig10} and \ref{fig:fig11} (for the cut-off criteria $I$ and $IV$ respectively), plotted in Cartesian coordinate system (X,Y) to understand the variation of metallicity in the plane of the sky. The number of regions with poor fit (mostly with slope lower than 3 in Figure \ref{fig:fig08}) decreases as one moves from criteria $I$ to $IV$. The number of such regions are only a few tens. Also, the metallicity maps using cut-off criteria II and III appear very similar and do not add much value to our results.  Thus, we prefer to present only the extreme cases, one with the relaxed cut-off ($I$) and one with the most stringent cut-off ($IV$) to bring out the differences in our results.
The value of the SMC centre used is, RA = 0$^h$ 52$^m$ 12.5$^s$; Dec = $-$72$^{\circ}$ 49$^m$ 43$^s$ (J2000.0 \cite{deVaucouleurs&Freeman1972VAstructure}).

The metallicity trend in the central SMC, eastern and western wings are seen in the OGLE III metallicity maps. The central region is metal rich but not homogeneous. There seems to be a very shallow gradient as one goes out from the centre towards the eastern or western regions. As we move from criteria $I$ to $IV$, more regions with poor fit are excluded from the map but the overall appearance hardly varies. The regions removed due to poor slope estimation are located in the centre and northeast, and are primarily the star forming regions. Such regions suffer from problem due to differential reddening and/or multiple-dominant population. The sub-regions with variation in LOS depth in the central and northeast part of the SMC that contribute to poor RGB slope-estimation also get removed in our analysis using the cut-off criteria. In both the maps the metallicity mainly varies from $-$0.8 dex to about $-$1.1 dex, with only a very few points more metal rich than $-$0.8 dex (mostly located near the central region) or more metal poor than $-$1.1 dex (mostly located away from the central region). 

\begin{figure*}
\centering
\begin{minipage}[b]{0.45\linewidth}
\includegraphics[height=3.0in,width=3.0in]{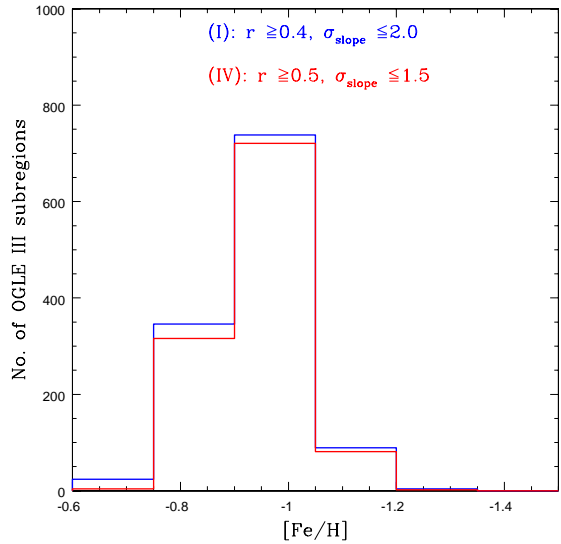}
\vskip 2cm
\caption{Histogram of $[Fe/H$] for OGLE III data, estimated for cut-off criteria $(I)$ in blue and $(IV)$ in red. $N_p$ $\ge$ 10 for both case.}
\label{fig:fig12} 
\end{minipage}
\end{figure*}

Figure \ref{fig:fig12} shows a histogram of metallicity for cut-off criteria $I$ and $IV$. The distribution is binned with a width of 0.15 dex, which is of the order of one sigma error. The distribution contains smaller number of regions with high metallicity as we progressively tighten the selection criteria, from $(I)$ to $(IV)$. The distribution peaks at about $-$0.95 dex, with a secondary peak at about $-$0.85 dex. There are hardly any regions with metallicity higher than $-$0.75 dex or lower than $-$1.2 dex. Table \ref{table:tab3} gives the list of mean metallicities of the SMC by region (along with their respective standard deviations) estimated for cut-off criteria $I$ and $IV$. The errors mentioned alongside the mean values are the standard deviation of the average, and do not include the error in metallicity estimation of each region. The mean values estimated by criteria $I$ and $IV$ are similar, though the number of regions considered changes. Nevertheless, criteria $IV$ being the most stringent one gives us the mean metallicity of the SMC. However, this value of metallicity estimate excludes the northern and southern parts of the SMC. To analyse those areas, we use MCPS data, discussed in the following section.


\begin{table*}
{\small
\caption{Mean metallicity for SMC using OGLE III data:}
\label{table:tab3}
\begin{tabular}{|c|c|c|c|c|c|c|}
\hline \hline
Cut-off criteria & $r$  & $\sigma_{slope}$  & Region of the SMC & Number of subregions & Mean [Fe/H] (dex) \\
\hline\hline
I   & $\ge$ 0.40 & $\le$ 2.0 & COMPLETE & 1201 & $-$0.94$\pm$0.09\\      
\hline    
IV  & $\ge$ 0.50 & $\le$ 1.5 & COMPLETE & 1124 & $-$0.94$\pm$0.08\\      

\hline       
\end{tabular}
\begin{minipage} {180mm}
\vskip 1.0ex
{Note: The first column denotes the first and fourth cut-off criteria considered to filter out the SMC subregions. It is to be noted that we considered $N_p$ $\ge$ 10 for both these cases. The second and third column specify the constraint on correlation coefficient ($r$) and $\sigma_{slope}$ respectively, corresponding to each cut-off criteria. The fourth column denotes complete coverage for OGLE III data. The number of subregions that satisfy the cut-off criteria, are mentioned in the fifth column. The mean metallicity and standard deviation for each cases are mentioned in the last (sixth) column.}
\end{minipage}
}
\end{table*}


\begin{table*}
{\small
\caption{Sub-division of MPCS regions:}
\label{table:tab4}
\begin{tabular}{|c|c|c|c|c|c|c|c|}
\hline \hline
Sl. no. & No. of  & No. of  & No. of     & No. of     & No. of             & Area of        &  Number of       \\
        & stars   & regions & division   & division   & sub-divisions      & a sub-division &  subregions      \\
        &         & (a)     & along RA   & along Dec  & (d=b$\times$c)     & (arcmin sq.)   &  (a$\times$d)    \\
        &         &         & (b)        & (c)        &                    &                &                  \\
\hline\hline
1  & 0 $<$ N $\le$ 2200      & 384 & 1 & 1 & 1  & (8.90$\times$10.0) & 384 (black)     \\
2  & 2200 $<$ N $\le$ 4200   & 148 & 2 & 1 & 2  & (4.45$\times$10.0) & 294 (brown)    \\
3  & 4200 $<$ N $\le$ 6000   & 62  & 3 & 1 & 3  & (2.97$\times$10.0) & 186 (red)       \\
4  & 6000 $<$ N $\le$ 8000   & 54  & 2 & 2 & 4  & (4.45$\times$5.0) & 216 (orange)    \\
5  &          N $>$ 8000     & 46  & 3 & 2 & 6  & (2.97$\times$5.0) & 276 (yellow)    \\
\hline     
\end{tabular}
\vskip 1.0ex
\begin{minipage} {180mm}
{Note: The table describes the 5 binning criteria used to sub-divide MCPS regions. The column descriptions are same as Table \ref{table:tab1}. The colours adjacent to the numbers in the last column are used to denote them in Figures \ref{fig:fig15} and \ref{fig:fig16}.}
\end{minipage}
}
\end{table*} 
\begin{figure*}
\centering
\begin{minipage}[b]{0.45\linewidth}
\includegraphics[height=3.0in,width=3.0in]{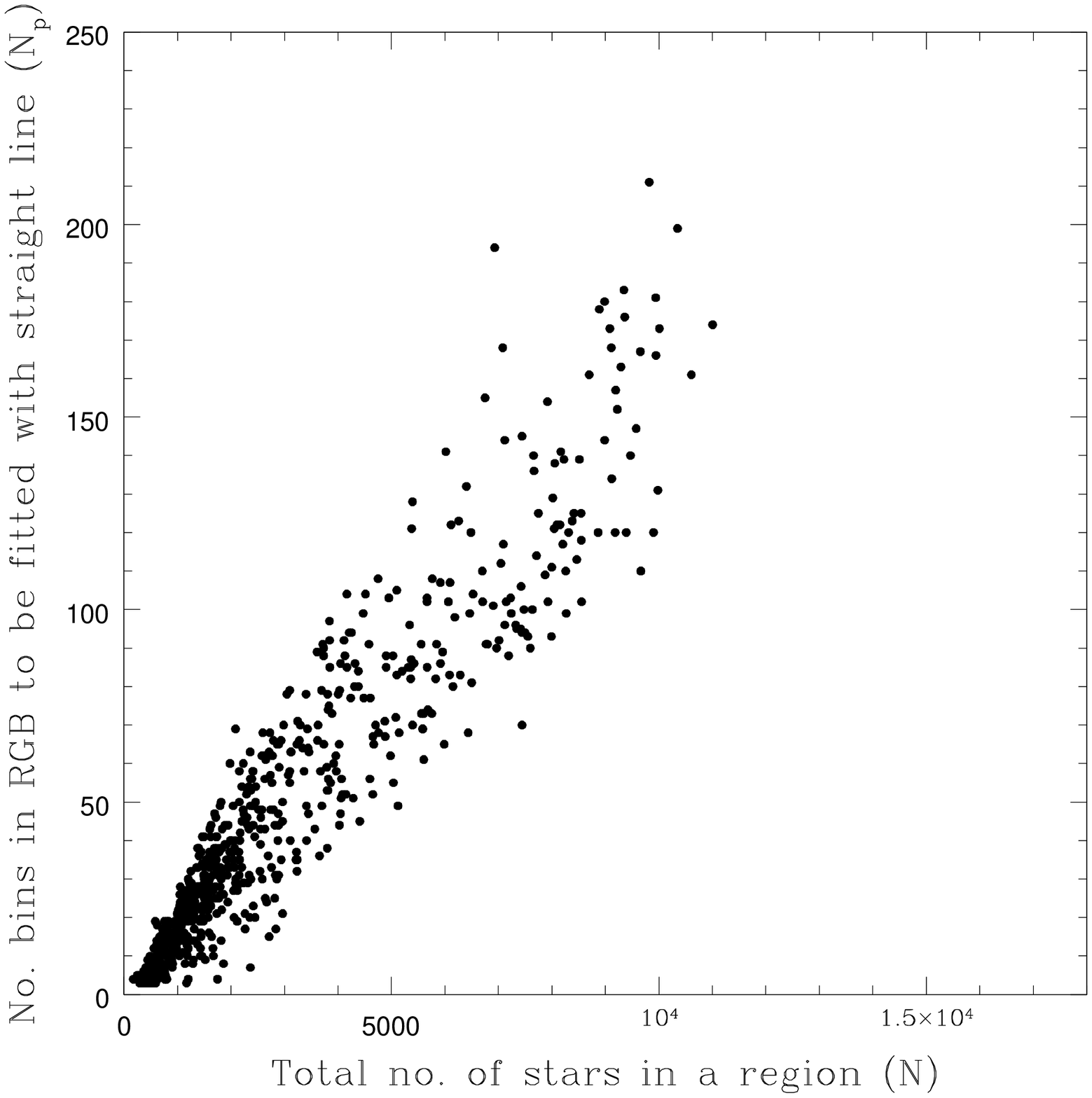}
  \vskip 2cm
  \caption{Plot of number of bins in RGB to be fitted with straight line ($N_p$) versus the total number of stars ($N$) for MCPS subregions, after initial area binning}
  \label{fig:fig13}
\end{minipage}
\quad
\begin{minipage}[b]{0.45\linewidth}
\includegraphics[height=3.0in,width=3.0in]{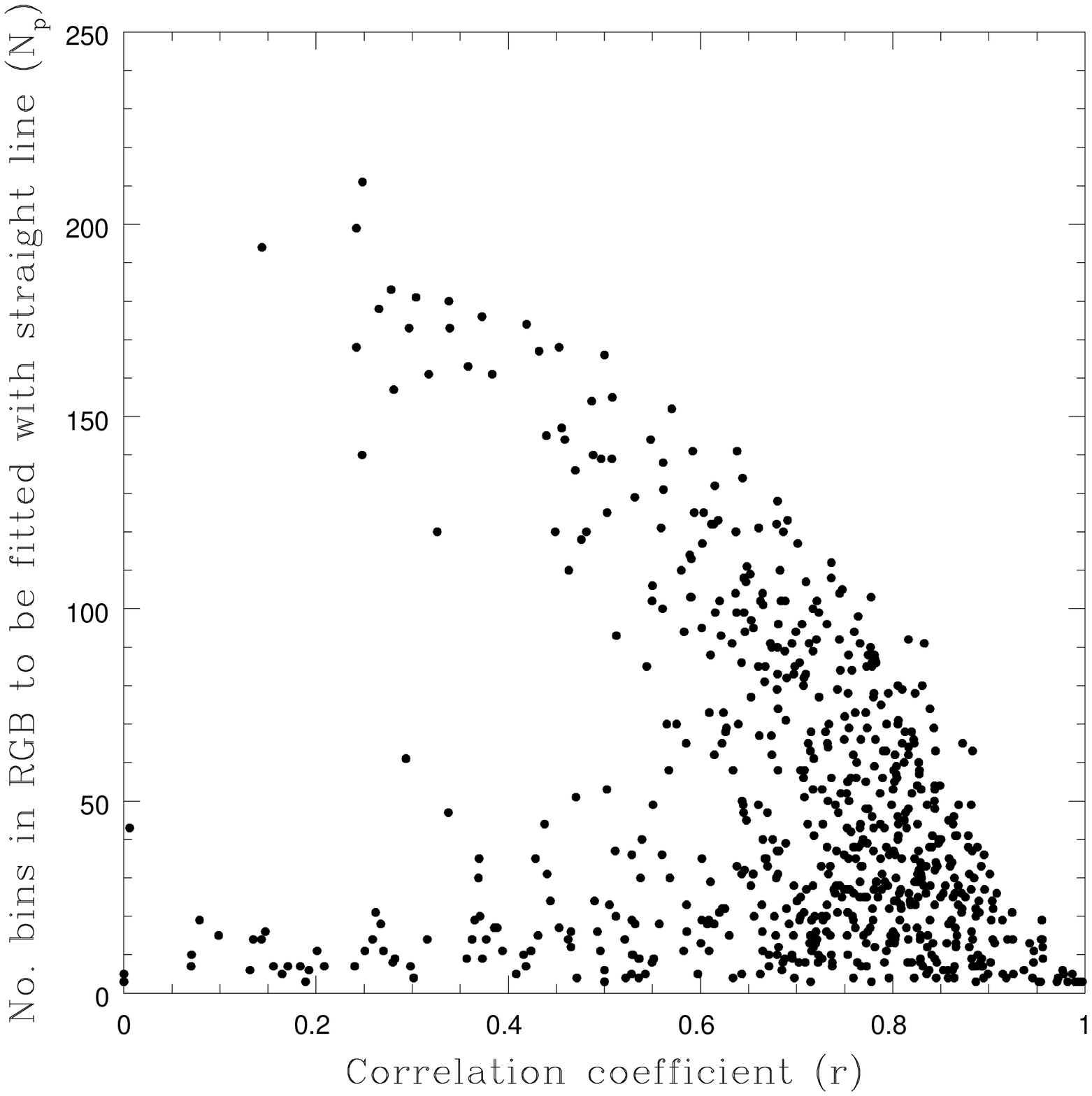}
  \vskip 2cm
  \caption{\small Plot of  number of bins in RGB to be fitted with straight line ($N_p$) versus correlation coefficient ($r$) for MCPS subregions, after initial area binning.}
  \label{fig:fig14}
\end{minipage}
\end{figure*}

\begin{figure*}
\centering
\begin{minipage}[b]{0.45\linewidth}
\includegraphics[height=3.0in,width=3.0in]{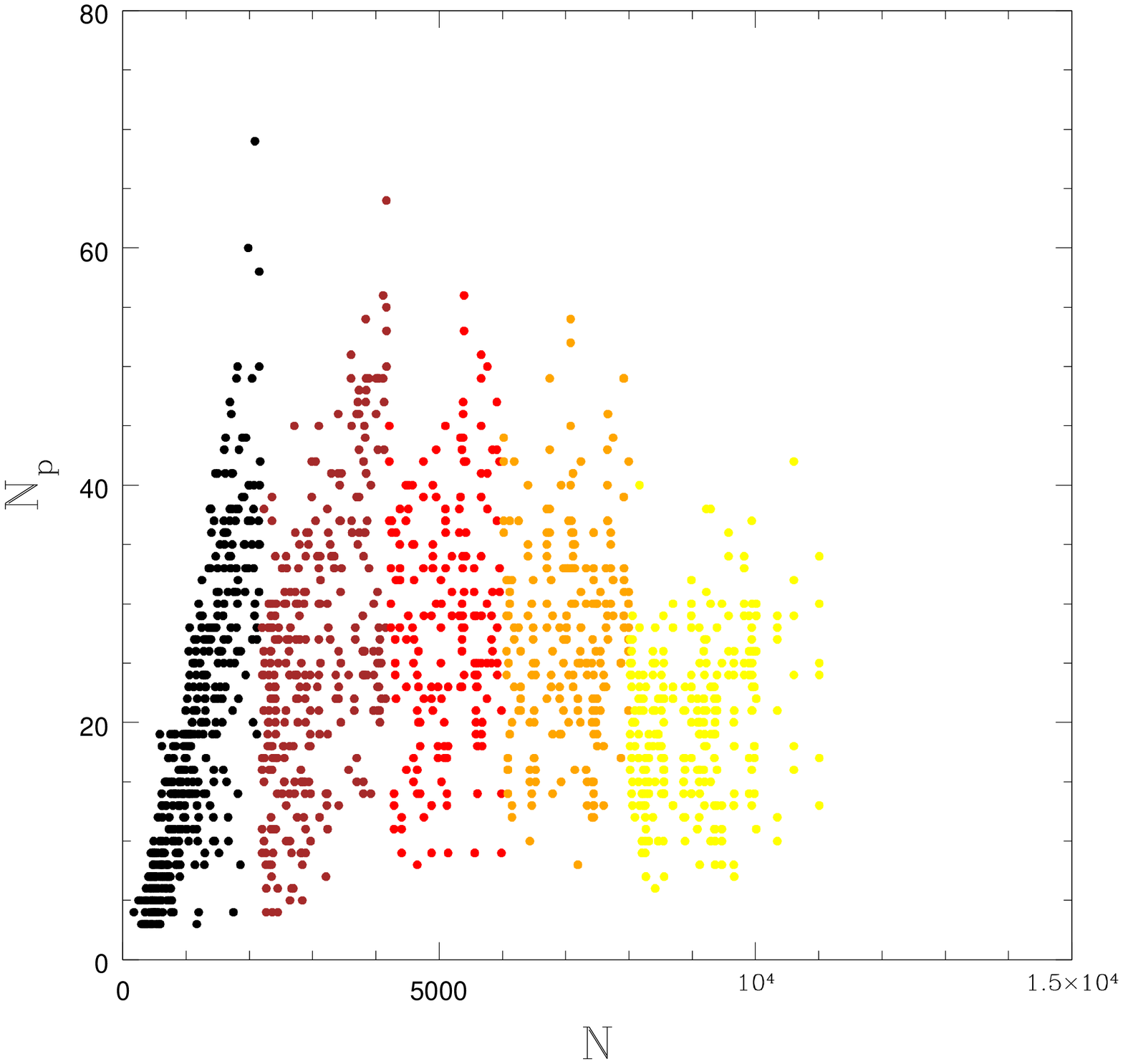}
  \vskip 2cm
  \caption{Plot of $N_p$ versus $N$ for MCPS subregions, after finer area binning. The colours correspond to the five different bin areas, as mentioned in the eighth column of Table \ref{table:tab4}.}
  \label{fig:fig15}
\end{minipage}
\quad
\begin{minipage}[b]{0.45\linewidth}
\includegraphics[height=3.0in,width=3.0in]{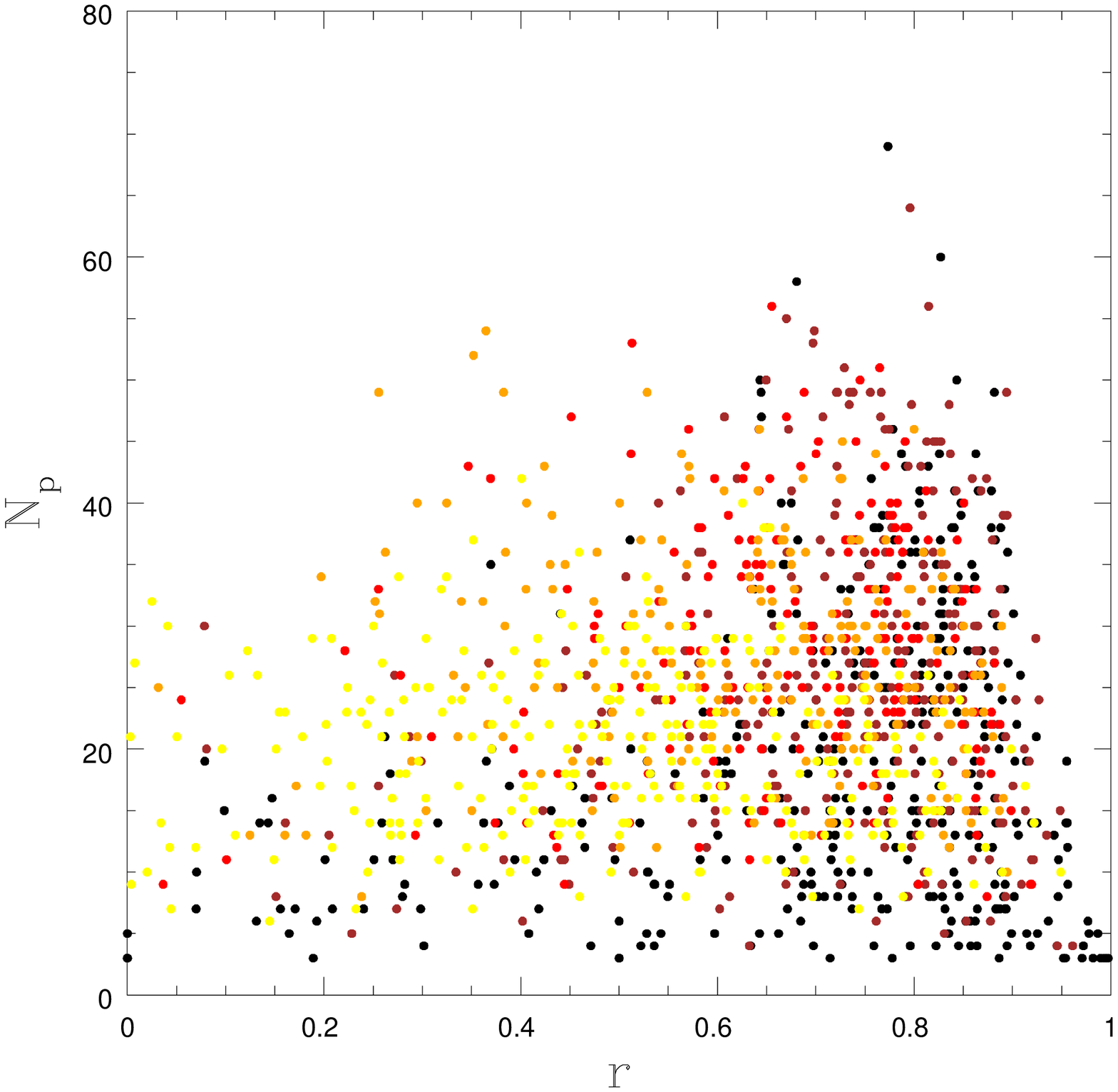}
  \vskip 2cm
  \caption{Plot of $N_p$ versus $r$ for MCPS subregions, after finer area binning. The colours correspond to the five different bin areas, as mentioned in the eighth column of Table \ref{table:tab4}.}
  \label{fig:fig16}
\end{minipage}
\end{figure*}
\begin{figure*}
\centering
\begin{minipage}[b]{0.45\linewidth}
\includegraphics[height=3.0in,width=3.0in]{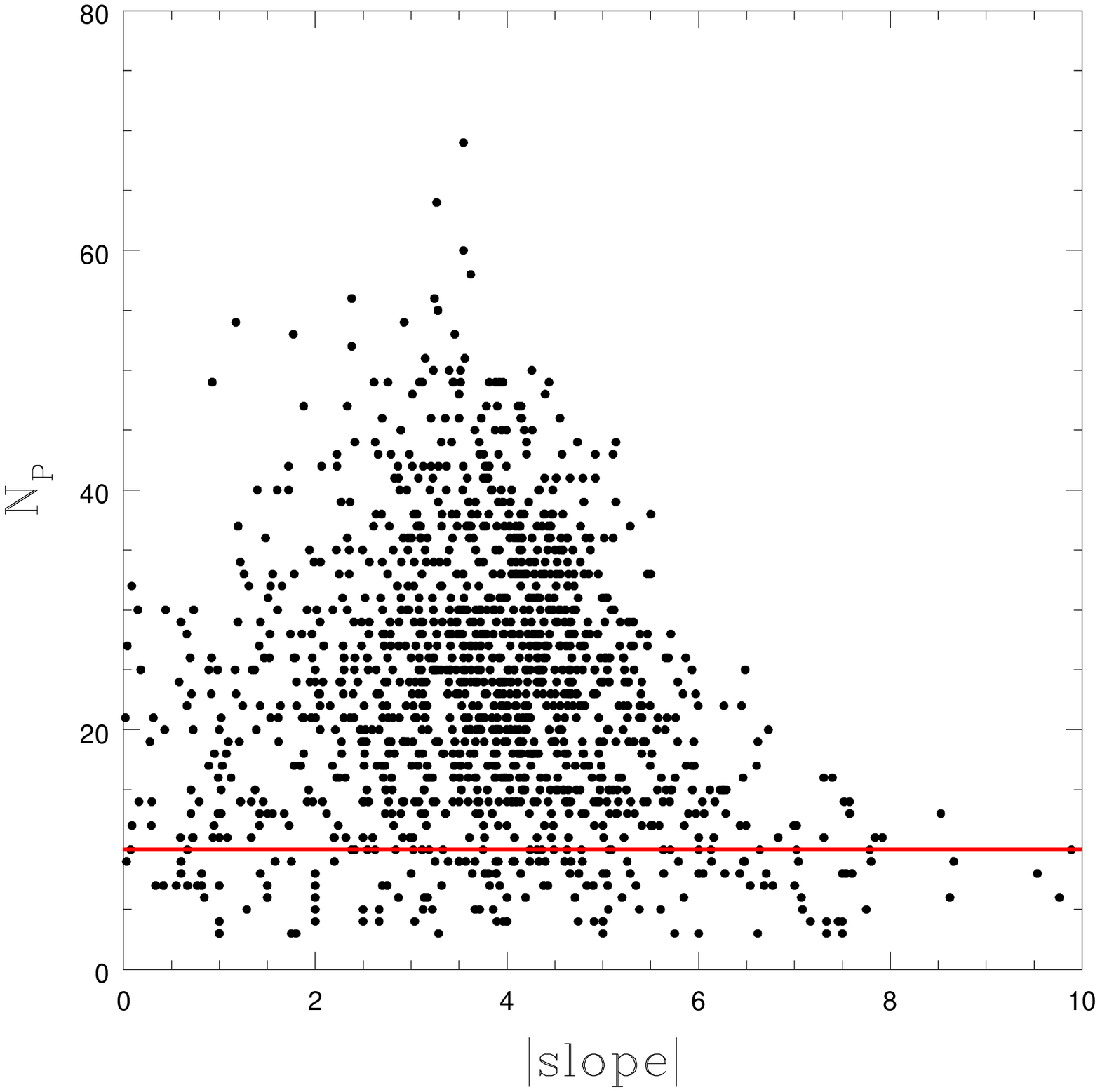}
  \vskip 2cm
  \caption{Plot of $N_p$ versus $|$slope$|$ for MCPS subregions. The red line at $N_p$ = 10 denotes the cut-off decided to exclude regions with poorly populated RGB.}
  \label{fig:fig17}
\end{minipage}
\quad
\begin{minipage}[b]{0.45\linewidth}
\includegraphics[height=3.0in,width=3.0in]{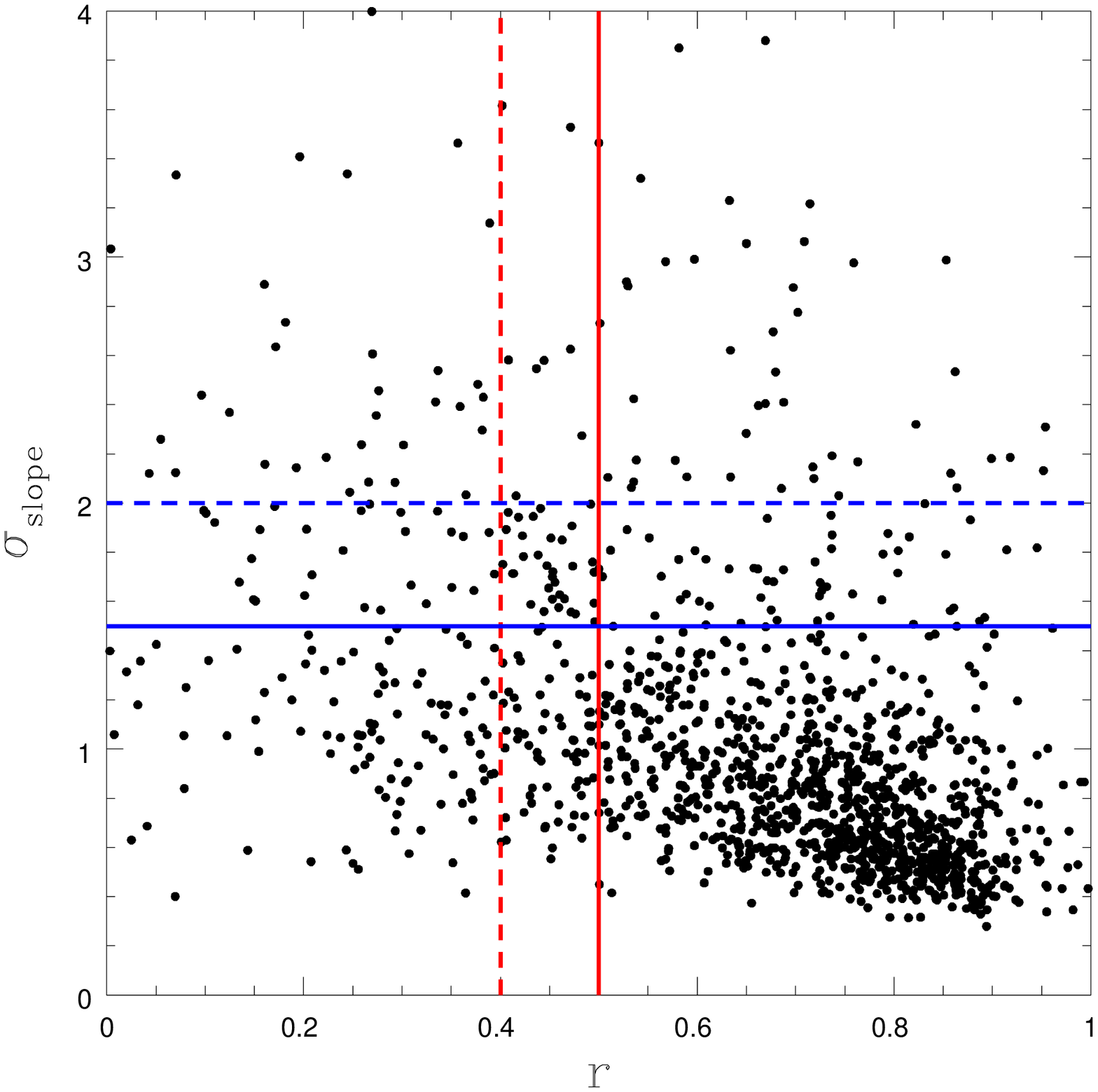}
  \vskip 2cm
  \caption{Plot of $\sigma_{slope}$ versus $r$ for MCPS subregions. The blue dashed and solid lines corresponds to the cut-off criteria on $\sigma_{slope}$ at 2.0 and 1.5 respectively. The red dashed and solid lines denote the cut-off corresponding to $r$ at 0.4 and 0.5 respectively.}
  \label{fig:fig18}
\end{minipage}
\end{figure*} 

\begin{figure*} 
\begin{center} 
\includegraphics[height=4.0in,width=4.0in]{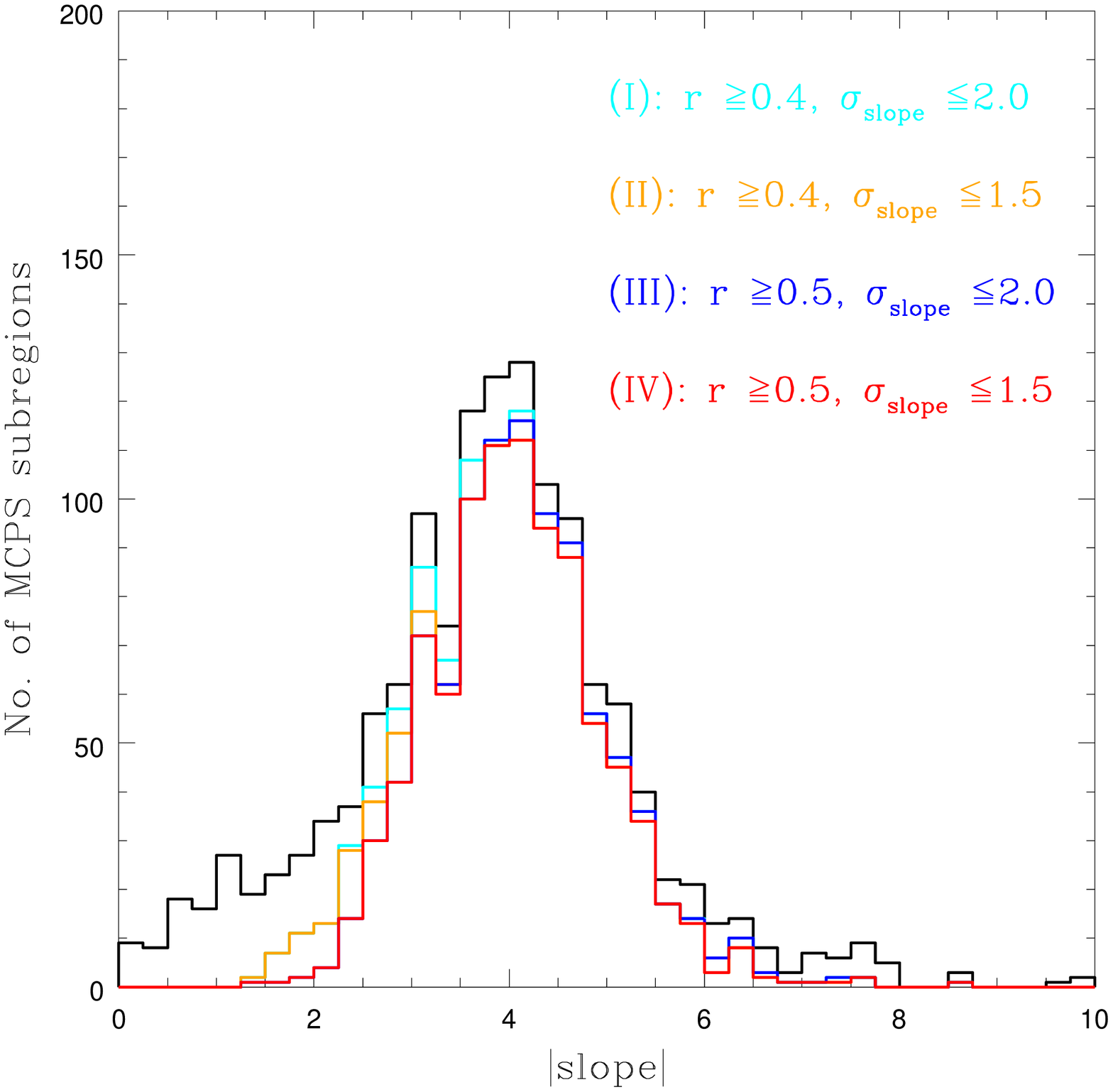}
  \vskip 2cm
  \caption{Histogram of $|$slope$|$ for MCPS subregions estimated for all the four cut-off criteria ($(I)$ in cyan, $(II)$ in orange, $(III)$ in blue, and $(IV)$ in red). $N_p$ $\ge$ 10 for all these four cases. The black solid line shows the distribution of $|$slope$|$ with no cut-off corresponding to all MCPS subregions.}
  \label{fig:fig19}
\end{center} 
\end{figure*}

\section{MCPS Analysis}

A similar analysis was carried out for the MCPS data, where the observed region is binned into 778 small sub-regions, each of dimension (8.9 $\times$ 10.0) sq. arcmin in RA and Dec. As already mentioned in Section 2, this survey covers a larger area than the OGLE III survey. Although its resolution is comparatively lower than the OGLE III survey, the MCPS provides more coverage on the northern and southern regions of the central SMC. We adopt the same procedure to estimate the slope of the RGB as was done for OGLE III analysis, starting from isolating the RGB, locating the densest point of the RC, and estimating the density distribution of stars on the RGB, and then estimating the RGB slope. The denser sub-regions were further subdivided based solely on stellar density. We adopted 5 binning criteria based on stellar density for the MCPS observed region as shown in Table \ref{table:tab4}. 

There are 1356 sub-regions analysed after finer area binning for which the parameters are estimated. The largest bins are (8$\farcm$90$\times$10$\farcm$0), and the smallest bins are (2$\farcm$97$\times$5$\farcm$0). Figures \ref{fig:fig13} and \ref{fig:fig14} show the plot of $N_p$ versus N and $N_p$ versus $r$ respectively, before finer area binning. For comparison, Figures \ref{fig:fig15} and \ref{fig:fig16} show the plot of $N_p$ versus N and and $N_p$ versus $r$ respectively, after finer area binning. Similar to the OGLE III analysis, while finely binning the observed area spatially, we have tried to constrain $N_p$ to similar values for all sub-regions in order to attain higher $r$ ($\ge$0.5). The regions with lower values of $r$ ($<$ 0.5) are likely to be those that suffer either from issues of multiple dominant population, small scale variation in reddening, and/or variation in LOS depth.

Similar to OGLE III analysis (Section 3), we use four cut-off criteria shown in Figure \ref{fig:fig17} ($N_p$ and $|slope|$) and Figure \ref{fig:fig18} ($\sigma_{slope}$ versus $r$) to filter out regions with poor slope estimations. The cut-off values for $r$ and $\sigma_{slope}$ (with $N_p$ $\ge$ 10 for all cases) are same as that were used for OGLE III. The slope distribution for all the four cut-off criteria are shown in Figure \ref{fig:fig19}, along with the distribution with no cut-offs. The figure shows that, as the $r$ cut-off becomes more stringent, the width of the histogram decreases such that the lower slope values are removed. The effect of $\sigma_{slope}$ is insignificant on the distribution for same cut-off in $r$. The strictest cut-off, $N_p$ $\ge$ 10, $r$ $\ge$ 0.5, and $\sigma_{slope}$ $\le$ 1.5 shows that slope values ranges primarily from 2 to 6 for the regions studied here. 

Although both OGLE III and MCPS deal with same optical bands (V and I filters), there may be systematic differences between their filter systems. In Paper I (their Equation 2), we found that there exists a systematic difference between their I bands (not in the V band). Therefore, we cannot directly compare the RGB slope distributions estimated from these two data sets. Thus, it is required to establish an independent slope-metallicity relation for the MCPS data, so as to bring the two surveys on to the same scale in metallicity.
\begin{table*}
{\small
\caption{Calibrators for MCPS slope-metallicity relation:}
\label{table:tab5}
\begin{tabular}{|c|c|c|c|c|c|c|c|c}
\hline \hline
RA$^{\circ}$ & Dec$^{\circ}$ & $|$slope$|$ & $\sigma_{slope}$ & $r$ & Mean [Fe/H] & Standard error \\
             &               &             &                  &     &    (dex)    &  of mean [Fe/H]\\
             &               &             &                  &     &             &                \\
\hline \hline

      8.26 &  -72.32  &   4.79  &    0.70  &    0.81  &  -1.04  &    0.07  \\
      9.25 &  -72.15  &   4.92  &    0.51  &    0.84  &  -1.10  &    0.07  \\
      9.75 &  -71.99  &   4.87  &    0.42  &    0.89  &  -1.08  &    0.07  \\
      9.75 &  -72.15  &   4.39  &    0.52  &    0.80  &  -0.87  &    0.07  \\
      9.76 &  -72.32  &   4.73  &    0.43  &    0.86  &  -1.00  &    0.05  \\
     10.25 &  -71.82  &   5.22  &    0.55  &    0.90  &  -0.98  &    0.07  \\
     10.26 &  -71.99  &   3.71  &    0.45  &    0.79  &  -0.93  &    0.06  \\
     10.26 &  -72.15  &   4.87  &    0.59  &    0.82  &  -1.01  &    0.06  \\
     10.75 &  -71.65  &   4.68  &    0.70  &    0.82  &  -0.88  &    0.07  \\
     10.76 &  -71.82  &   4.82  &    0.57  &    0.85  &  -1.01  &    0.07  \\
     10.75 &  -71.99  &   4.99  &    0.50  &    0.88  &  -1.02  &    0.07  \\
     18.74 &  -72.15  &   5.99  &    0.75  &    0.87  &  -1.07  &    0.06  \\
     18.74 &  -72.32  &   4.58  &    0.62  &    0.79  &  -1.03  &    0.05  \\
      8.76 &  -72.48  &   4.58  &    0.48  &    0.86  &  -0.92  &    0.06  \\
      7.75 &  -74.14  &   4.09  &    0.46  &    0.84  &  -0.97  &    0.07  \\
      8.75 &  -74.14  &   3.60  &    0.48  &    0.77  &  -1.08  &    0.07  \\
     15.25 &  -74.15  &   4.26  &    0.39  &    0.84  &  -1.18  &    0.07  \\
     19.75 &  -73.65  &   3.91  &    0.87  &    0.73  &  -1.10  &    0.07  \\
     10.75 &  -74.65  &   3.68  &    0.65  &    0.74  &  -1.12  &    0.07  \\
     13.25 &  -74.48  &   4.44  &    0.35  &    0.88  &  -1.03  &    0.06  \\
     13.75 &  -74.48  &   5.14  &    0.54  &    0.82  &  -1.20  &    0.06  \\
     13.74 &  -74.65  &   4.95  &    0.53  &    0.86  &  -1.05  &    0.06  \\
     15.24 &  -74.81  &   6.04  &    0.82  &    0.90  &  -1.22  &    0.06  \\
     11.88 &  -72.15  &   4.26  &    0.44  &    0.83  &  -0.85  &    0.07  \\
     12.88 &  -71.99  &   3.96  &    0.55  &    0.72  &  -0.90  &    0.07  \\
     13.13 &  -71.82  &   4.36  &    0.38  &    0.93  &  -1.05  &    0.06  \\
     14.88 &  -71.65  &   4.73  &    0.63  &    0.84  &  -0.93  &    0.05  \\
     16.87 &  -71.99  &   4.12  &    0.80  &    0.76  &  -0.78  &    0.07  \\
      8.37 &  -72.99  &   3.86  &    0.44  &    0.86  &  -0.91  &    0.07  \\
     10.38 &  -72.49  &   4.51  &    0.41  &    0.87  &  -1.00  &    0.07  \\
     18.12 &  -72.65  &   4.48  &    0.52  &    0.88  &  -0.96  &    0.07  \\
     10.87 &  -73.82  &   3.80  &    0.39  &    0.84  &  -0.80  &    0.07  \\
     13.25 &  -71.99  &   4.67  &    0.66  &    0.80  &  -0.80  &    0.07  \\
     16.75 &  -72.48  &   3.96  &    0.63  &    0.76  &  -0.97  &    0.07  \\
     16.75 &  -72.65  &   3.66  &    0.62  &    0.78  &  -0.91  &    0.06  \\
     12.25 &  -73.65  &   3.40  &    0.44  &    0.74  &  -0.84  &    0.07  \\
     10.37 &  -72.86  &   3.37  &    0.51  &    0.72  &  -0.85  &    0.07  \\
     10.13 &  -73.11  &   4.94  &    0.63  &    0.84  &  -1.09  &    0.07  \\
     16.62 &  -72.77  &   4.52  &    0.89  &    0.76  &  -0.90  &    0.07  \\
     11.87 &  -73.44  &   2.89  &    0.42  &    0.73  &  -0.85  &    0.06  \\
     12.63 &  -73.52  &   4.60  &    0.87  &    0.71  &  -0.92  &    0.07  \\

\hline     
\end{tabular}
\begin{minipage} {180mm}
\vskip 1.0ex
{Note: The table lists out the 41 calibrators used to construct the slope-metallicity relation for MCPS data. The column descriptions are same as Table \ref{table:tab2}.}
\end{minipage}
}
\end{table*} 
\begin{figure*} 
\begin{center} 
\includegraphics[height=4in,width=4in]{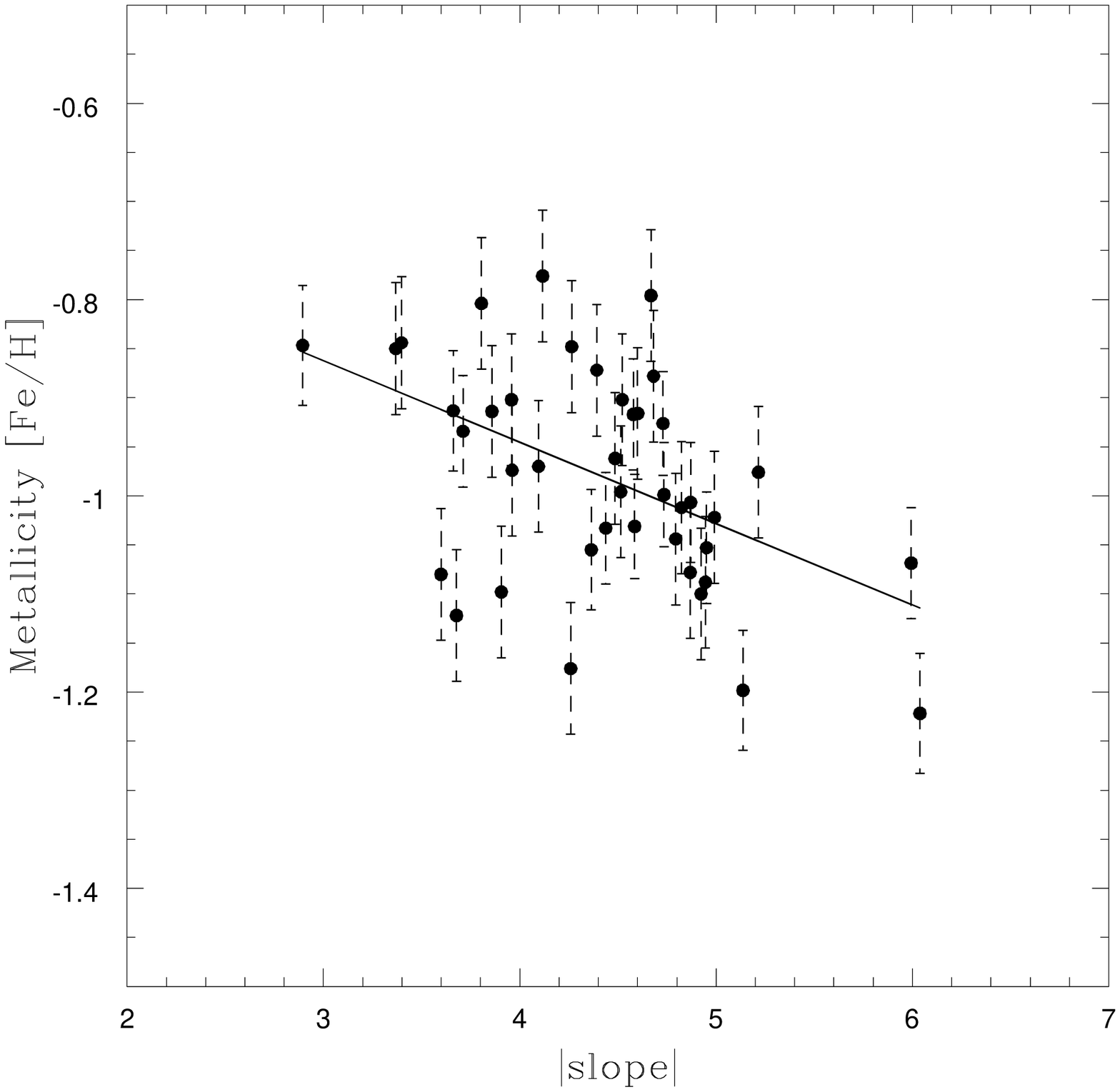}
\vskip 2cm
\caption{Plot of [Fe/H] versus $|$slope$|$ for MCPS data. The points denote our subregions whose mean [Fe/H] has been found using RGs from \citealt{Dobbie+2014MNRAS-papII}, with the solid line denoting a linear relation between them. The error bar (dashed line) shown for each point is the standard error of mean [Fe/H].}
\label{fig:fig20}
\end{center} 
\end{figure*}

\begin{figure*} 
\begin{center} 
\includegraphics[height=5in,width=7in]{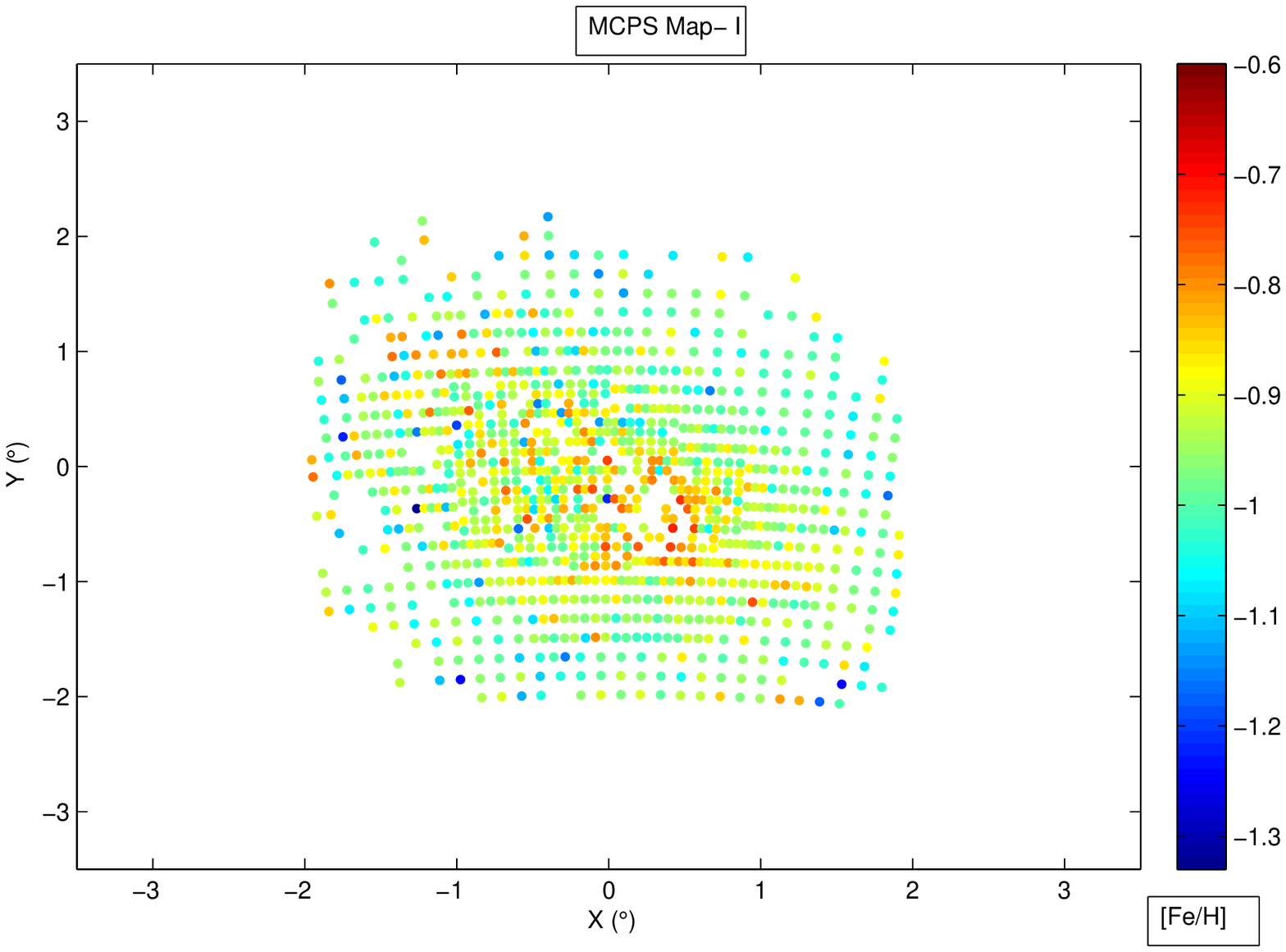}
\vskip 6cm
\caption{MCPS metallicity map with cut-off criteria $(I)$: $N_p$ $\ge$ 10, $r$ $\ge$ 0.4 and $\sigma_{slope}$ $\le$ 2.0.}
\label{fig:fig21}
\end{center} 
\end{figure*}

\begin{figure*} 
\begin{center} 
\includegraphics[height=5in,width=7in]{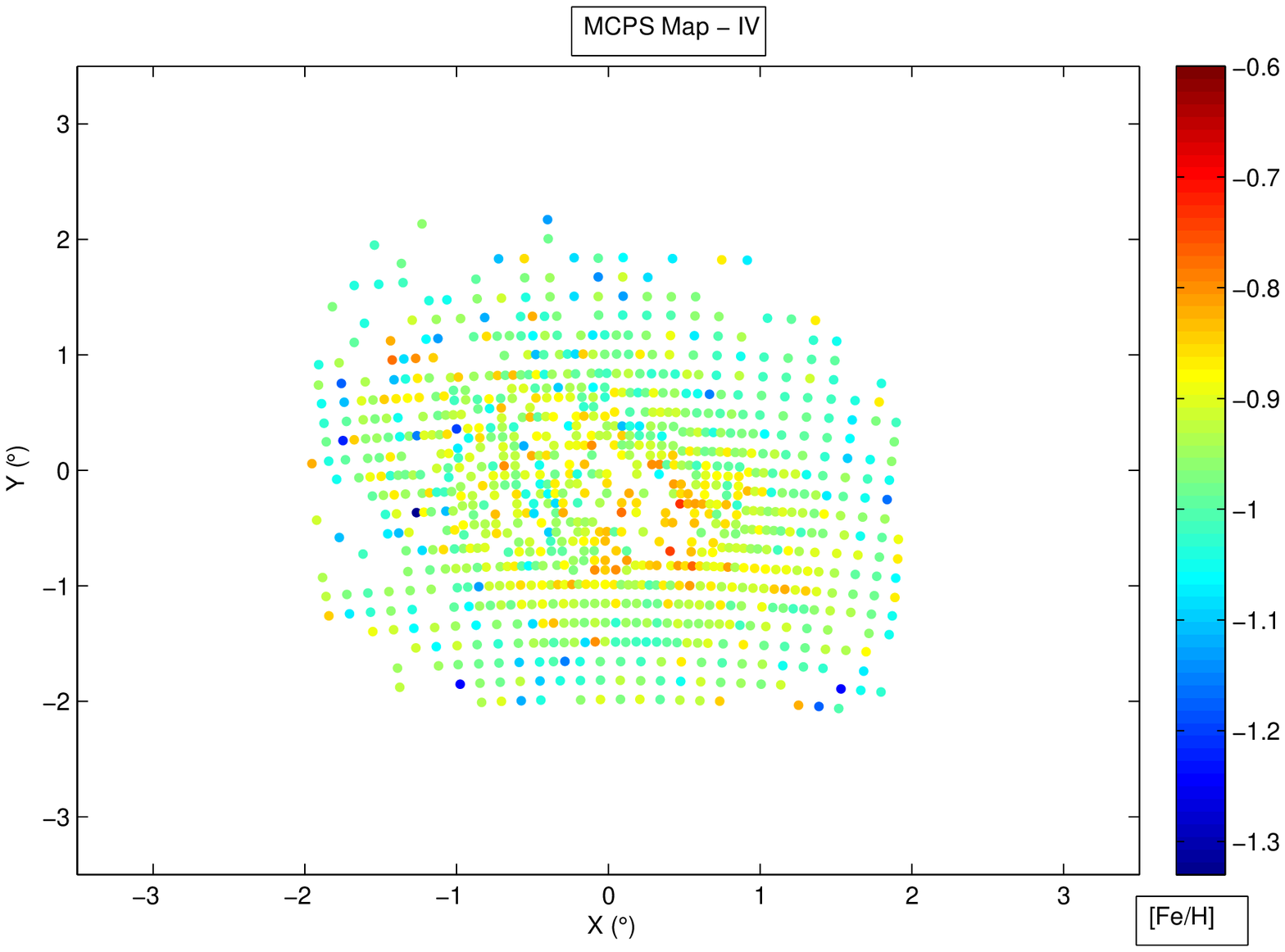}
\vskip 6cm
\caption{MCPS metallicity map with cut-off criteria $(IV)$: $N_p$ $\ge$ 10, $r$ $\ge$ 0.5 and $\sigma_{slope}$ $\le$ 1.5.}
\label{fig:fig22} 
\end{center} 
\end{figure*}


\subsection{Calibration of RGB slope to Metallicity}

We adopt similar technique as in Section 3.1 for OGLE III case to estimate a slope-metallicity relation for MCPS data set. Since MCPS has a larger spatial coverage than OGLE III, we are able to employ more data points to calibrate the RGB slopes. The metallicity and slope ranges obtained for the MCPS calibration points are similar to that for OGLE III data set. The slope range of the calibration points also more or less covers the entire slope distribution (Figure \ref{fig:fig19}) of the MCPS data. Figure \ref{fig:fig20} shows the MCPS slope metallicity relation derived by a linear fit after 2$\sigma$ clipping. The final calibration points (41) are listed down in Table \ref{table:tab5}.
\begin{equation} \label{eq:2}
[Fe/H]=(-0.083\pm 0.024)\times|slope|+ (-0.614\pm 0.106);
\end{equation}
with $r$=0.49. This is different from the calibration of MCPS metallicity map for the LMC (Section 4.2, Paper I). There, due to inadequate number of calibration points for the MCPS case, we transformed the MCPS filter system to the OGLE III filter system by correcting for their systematic difference between I filters. Then, we used the OGLE III slope metallicity relation to calibrate the MCPS slopes to metallicities. However, in this case we successfully established an independent slope-metallicity relation for MCPS data. Thus, the OGLE III and MCPS system are now tied down to the same scale in metallicity.

\begin{table*}
{\small
\caption{Mean metallicity for SMC using MCPS data:}
\label{table:tab6}
\begin{tabular}{|c|c|c|c|c|c|c|}
\hline \hline
Cut-off criteria & $r$  & $\sigma_{slope}$  & Region of the SMC & Number of subregions & Mean [Fe/H] (dex) \\
\hline\hline
I   & $\ge$ 0.40 & $\le$ 2.0 & COMPLETE & 1035 & $-$0.94$\pm$0.08\\      
\hline    
IV  & $\ge$ 0.50 & $\le$ 1.5 & COMPLETE & 913  & $-$0.95$\pm$0.07\\      

\hline       
\end{tabular}
\begin{minipage} {180mm}
\vskip 1.0ex
{Note: The column descriptions are same as Table \ref{table:tab3}, except that the fourth column denotes complete coverage for MCPS data.}
\end{minipage}
}
\end{table*}

\begin{figure*}
\centering
\begin{minipage}[b]{0.45\linewidth}
\includegraphics[height=3.0in,width=3.0in]{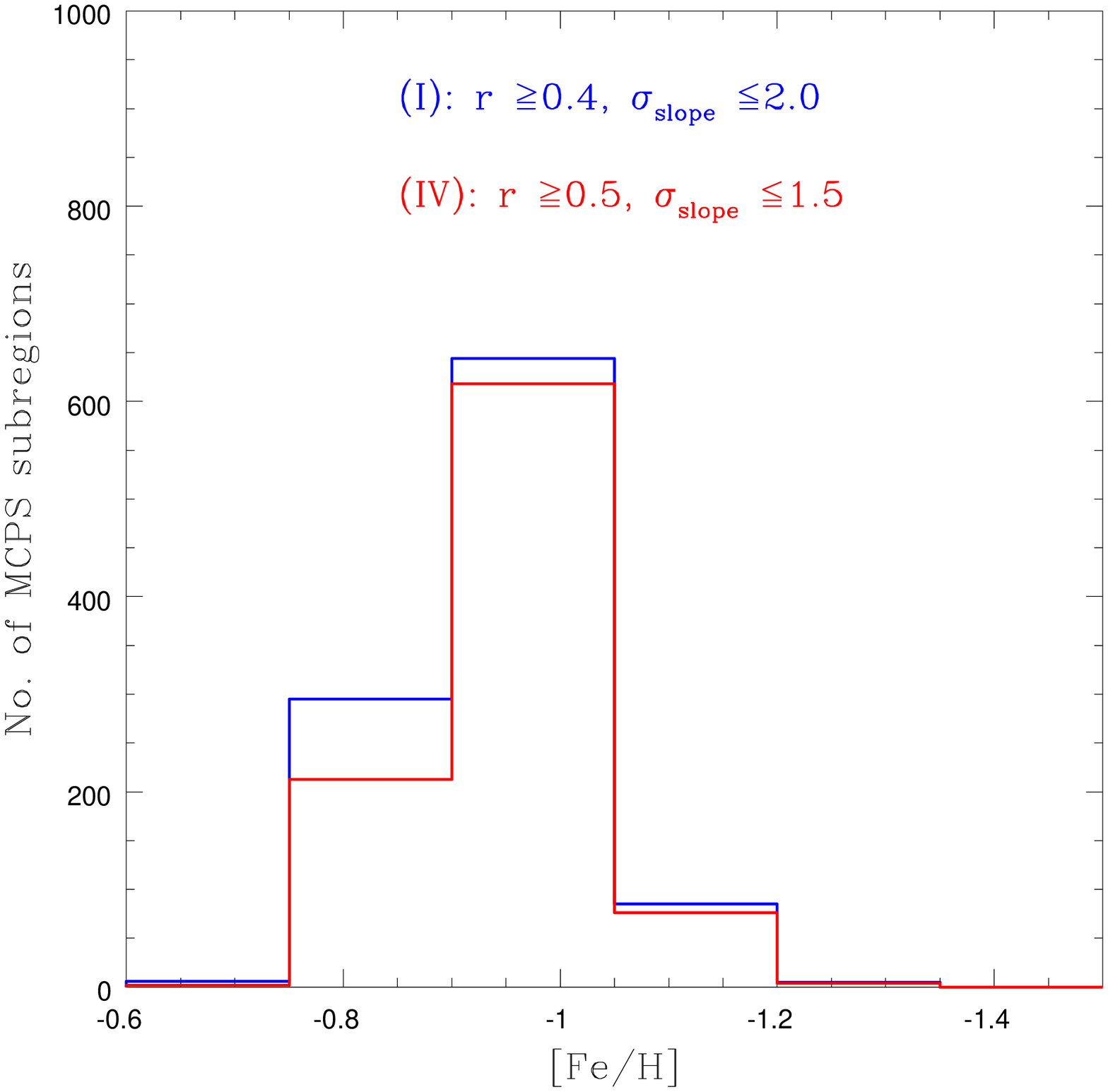}
\vskip 2cm
\caption{Histogram of $[Fe/H$] for MCPS data, estimated for cut-off criteria $(I)$ in blue and $(IV)$ in red. $N_p$ $\ge$ 10 for both case.}
\label{fig:fig23}
\end{minipage}
\end{figure*} 


\subsection{MCPS Metallicity Map}

The MCPS metallicity maps are shown in Figure \ref{fig:fig21} and \ref{fig:fig22} for the cut-off criteria $I$ and $IV$ respectively (using Equation \ref{eq:2}), in Cartesian coordinate (X,Y) system on the projected sky plane. For reasons similar to the OGLE III case we have ignored the maps using criteria $II$ and $III$ and presented only the two extreme cases. The metallicity trend in the northern and southern SMC are now revealed. A qualitative comparison shows that these regions have less metallicity variations as compared to the eastern and western SMC. The map shows a shallow gradient from central to outer SMC within 2.5 degrees. Again, more regions with poor fit are excluded as we move from cut-off criteria $I$ to $V$. The metallicity ranges from $-$0.8 dex to about $-$1.1 dex, with only very few points more metal rich than $-$0.8 dex (mostly located near the central region) or more metal poor than $-$1.1 dex (mostly located away from the central region). The gaps in the map are due to regions with poorly estimated slopes. The regions which get missed out due to this effect are those located near the central, northeast and star forming regions (similar to OGLE III). 

Figure \ref{fig:fig23} shows a histogram of metallicity for cut-off criteria $I$ and $IV$. The distribution is binned with a width of 0.15 dex. The distribution looks very similar to that of the OGLE III distribution (but with fewer subregions), and peaks at about $-$0.95 dex alongside a secondary peak at about $-$0.85 dex. Table \ref{table:tab6} lists the mean metallicity (along with standard deviation) of the SMC estimated for cut-off criteria $I$ and $IV$ using the MCPS data. The errors mentioned alongside the mean values are the standard deviation of the average, and does not include the error in metallicity estimation of each region. It is to be noted that the mean values for both the criteria are almost similar, though the number of regions changes. The mean values also correspond to that estimated using the OGLE III data in Table \ref{table:tab3}. Since, $IV$ is the most-stringent criteria, the rest of our analysis for both the data-sets will be based on that.


\section{Error Analysis}
We describe the error estimation of slope and metallicity for both the data sets in this section. The method is similar to that described in Paper I (Section 5). The factors that can contribute to the error in slope estimation are the photometric error associated with individual points in the CMD (error in V and I magnitude are $\le$ 0.15), and error due to fine binning of the CMD (dimension of each bin is 0.05 in  colour, and 0.10 in magnitude). In this study, we are interested in the mean metallicity of a region, of which the RGB slope is an indicator. The above factors could contribute to the errors if we were concerned with the probability of a star lying within a particular CMD bin, and the resulting density of stars within individual CMD bins while estimating the RGB slope. However, under the present scenario we are concentrating only on the overall RGB feature and not looking into the strength of individual CMD bins. After we identify the most populated bins as a part of the RGB, the spread in the distribution of RGB bins is such that it more or less makes a good representation of the RGB. We fit these bins with the standard least square fitting technique, to estimate the slope ($|$slope$|$) and its corresponding error ($\sigma_{slope}$). The values of $\sigma_{slope}$ estimated in our study are relatively larger, caused by the natural spread of the populated RGB bins. This is shown in Figure \ref{fig:fig07} and \ref{fig:fig18} for OGLE III and MCPS respectively. There we can see that for most of the regions, $\sigma_{slope}$ shows a clumpy distribution below 2. Thus, in this study we do not take into account the errors associated with individual stars and colour-magnitude binning when estimating the error in slope. We expect these contributions to be negligible.

The error in metallicity for individual sub-regions are calculated by propagation of errors applied to the slope-metallicity relation (Equation \ref{eq:1} and \ref{eq:2}) for OGLE III and MCPS individually. 
We can express the error associated with metallicity ($error_{[Fe/H]}$) for each subregion as:
{\small
\begin{equation} \label{eq:3}
error_{[Fe/H]}=\sqrt{{(b \times |slope|)}^2\times\left(\left(\frac{\sigma_b}{b}\right)^2+\left(\frac{\sigma_{slope}}{|slope|}\right)^2\right)+{\sigma_a}^2}.
\end{equation} 
}
where $b$ is the slope and $a$ is the y-intercept of the slope-metallicity relation, and $\sigma_b$ and $\sigma_a$ are the their respective errors. It is to be noted that while calculating $error_{[Fe/H]}$, we have not considered the error in metallicity associated with individual calibration points.

\begin{figure*}
\centering
\begin{minipage}[b]{0.45\linewidth}
\includegraphics[height=3.0in,width=3.0in]{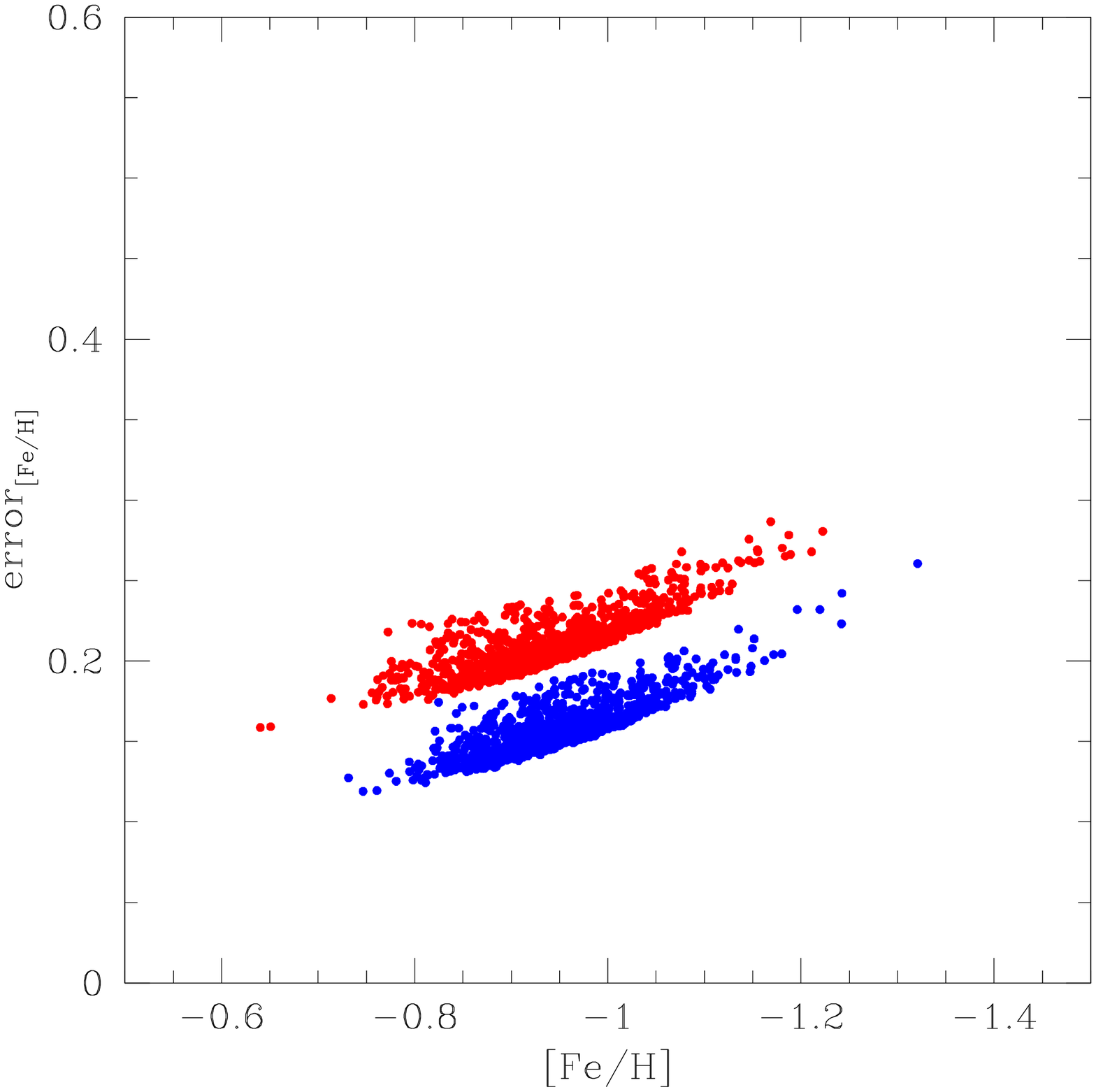}
  \vskip 2cm
  \caption{Plot of $error_{[Fe/H]}$ versus [Fe/H] for OGLE III (red filled circles) and MCPS (blue filled circles), for cut-off criteria $(IV)$.}
  \label{fig:fig24}
\end{minipage}
\quad
\begin{minipage}[b]{0.45\linewidth}
\includegraphics[height=3.0in,width=3.0in]{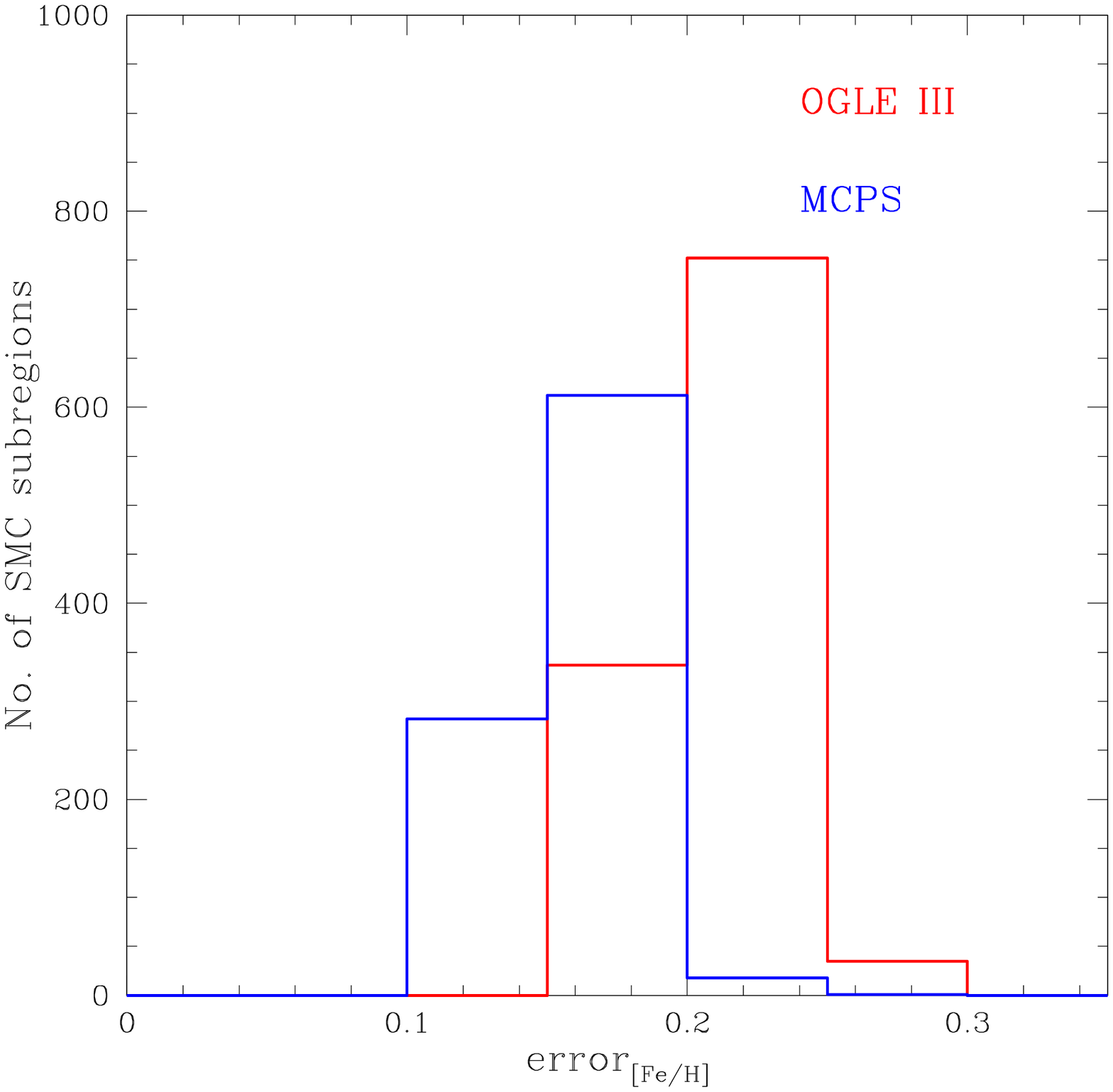}
  \vskip 2cm
  \caption{Plot of histogram of $error_{[Fe/H]}$ for OGLE III (red) and MCPS (blue), for cut-off criteria $IV$.}
  \label{fig:fig25}
\end{minipage}
\end{figure*}

The error in metallicity is plotted with respect to the [Fe/H] for cut-off criteria $IV$ for both data set and compared in the Figure \ref{fig:fig24}. As seen, the error for MCPS (blue) is in the range $\sim$ 0.10--0.25 whereas that for OGLE III (red) is in the range $\sim$ 0.15-0.30. Figure \ref{fig:fig25} shows the distribution of error for OGLE III and MCPS data. The distribution appears similar in both cases, except that the OGLE III distribution has more regions and is shifted with respect to the MCPS distribution by almost constant value of 0.05 dex.

\section{Discussion}
We have created maps of the photometrically-estimated metallicity of the SMC using OGLE III and MCPS data. The two data sets are complementary in terms of the area coverage of the SMC. The method traces the densest part of the RGB and the slope is estimated using a straight line fit. We thus estimate only the metallicity of the population which has the largest number of stars in the RGB phase. The analysis does not bring out any information regarding the contribution from the less dominant populations. We estimated the slope of the RGB of several subregions within the SMC using this method that takes care of reddening and density variation between regions. The RGB slope is then converted to metallicity using spectroscopic measures which nearly covers the full range. If there are multiple populations with similar number of stars, then the RGB tends to be broad and the slope may be poorly estimated. These regions are not considered in our analysis, due the removal of regions with poor estimation of slope. We discuss the assumptions made in this study and the implications of our results in the following sections. 

\subsection{Assumptions and their impact}
The steps undertaken to create the metallicity maps might have affected the outcome of the study. Below we discuss each step and its impact on the estimated value of RGB slope and metallicity. 

\subsubsection{Effect of sub-division}

The two data sets are spatially sub-divided to create regions with smaller area. We have also considered different sizes for sub-division based on the stellar density. As the depth and resolution of MCPS and OGLE III are different, we were unable to make the subregions the same area in both data sets. The MCPS subregions are relatively larger than the OGLE III subregions throughout the SMC. As a result, even though we bring them to the same metallicity scale, we are unable to make one-to one matches between the metallicity estimated using the two data sets. As the area of a subregion changes, the estimated slope changes mildly, though within the errors. This may be due to the fact that the dominant population as well as differential reddening can change with area.

\subsubsection{Effect of reddening and differential reddening}
In the reddening map presented by \cite{Smitha&Purni2012} using RC stars, the SMC is shown to have variation in reddening across the galaxy. Also, the south-west and north-east regions about the SMC centre and the Eastern Wing region have larger reddening compared to other regions within the SMC. The reddening variation can shift the location of the RGB in the CMD. This effect is taken care in the analysis by anchoring the RGB to the densest part of the RC. But, it is to be noted that the effect of differential reddening will remain and make the RGB broad. Large scale variation of reddening can broaden the RGB, resulting in a poorly estimated slope. Such regions are eliminated from our analysis. Most of the regions that get eliminated due to poor slope estimation, are found to be located near the regions with large variation in reddening (around the centre, and Eastern Wing). Since, we are interested in the statistical average estimated using large number of regions, their impact is likely to be negligible. 

\subsubsection{Effect of line of sight depth}
The SMC is known to have large LOS depth as traced by young, intermediate-age, and old stellar populations (\citealt{Mathewson+1986, Mathewson+1988, Crowl+2001, Smitha&Purni2012}). Recently \cite{Smitha+2017MNRAS} identified regions towards East of $\sim$ $-$2${^\circ}$ that show a foreground population ($\sim$11.8 $\pm$ 2.0 kpc in front of the main body) in the form of a distance bimodality in the RC distribution. The authors analyzed a larger area ($\sim$ 20 sq. degrees) than our observed region using near-infrared photometric data from the VISTA-VMC survey. They relate the bimodality in their extreme East regions to tidal stripping from the SMC during the most recent encounter with the LMC. The OGLE III observed field has a very few subregions ($<$ 4$\%$ of analysed subregions) beyond $\sim$ $-$2${^\circ}$ East, whereas, the MCPS has none. The tip of the RGB might vary in sharpness, being more spread out where the LOS depth is large, which may affect the estimation of slope by making the RGB appear steeper than it otherwise would for a given metallicity. Also, the RGB in these extreme East regions are sparsely populated as they lie at the periphery of the observed region. These factors can lead to poor estimation of RGB slope. Thus, the effect on our results by such subregions will be negligible. This is because, they either get excluded by cut-off criteria, or their percentage is insignificant compared to the total number of subregions analyzed. \cite{Smitha+2017MNRAS} also mention the presence of a very mild gradient in distance modulus from East to West of the SMC (between $-$2${^\circ}$ to 2${^\circ}$), and no gradient as such from North to South. However, according to \cite{Smitha&Purni2009, Smitha&Purni2012} who used the RC distribution from MCPS and OGLE (II and III) data to estimate the LOS depth of the SMC, it was found to that the LOS depth is almost uniform across the inner SMC. Thus, we are consistent with our method in estimating the slope of RGB and calibrating the same to metallicity within our observed area. In this study, we are interested in the mean feature of the RGB. So, for mild/uniform variation in depth within a region the effect of RGB spread on our result will be mild/unlikely. The authors point out that the regions around the SMC centre and north-east suffer from a larger LOS depth. These regions with large depth get removed from our map due to poor slope estimation using the cut-off criteria on the estimated parameters. 

\subsubsection{Systematics in the calibration}
There can be a systematic effect in our study related to the conversion of slope to metallicity. We have tried to minimise it by formulating independent slope-metallicity relation for OGLE III and MCPS data using the same spectroscopic study. Across the inner 2.5 degrees of the SMC, the RGB slope has a range of values; the calibration relation between slope and metallicity should hold good for the full range. We used the results of \cite{Dobbie+2014MNRAS-papII} which has a large range in metallicity to achieve this calibration independently for both the data set. The calibration of RGB slope to metallicity rests on the assumption that the spectroscopic targets are drawn from the dominant population of the subregion. This assumption is fair enough, as the spectroscopic targets are also chosen from the RGB and are likely to be picked from the population with most number of stars.

\subsection{Comparison of metallicity distribution}
 
\begin{figure*}
\centering
\begin{minipage}[b]{0.45\linewidth}
\includegraphics[height=3.0in,width=3.0in]{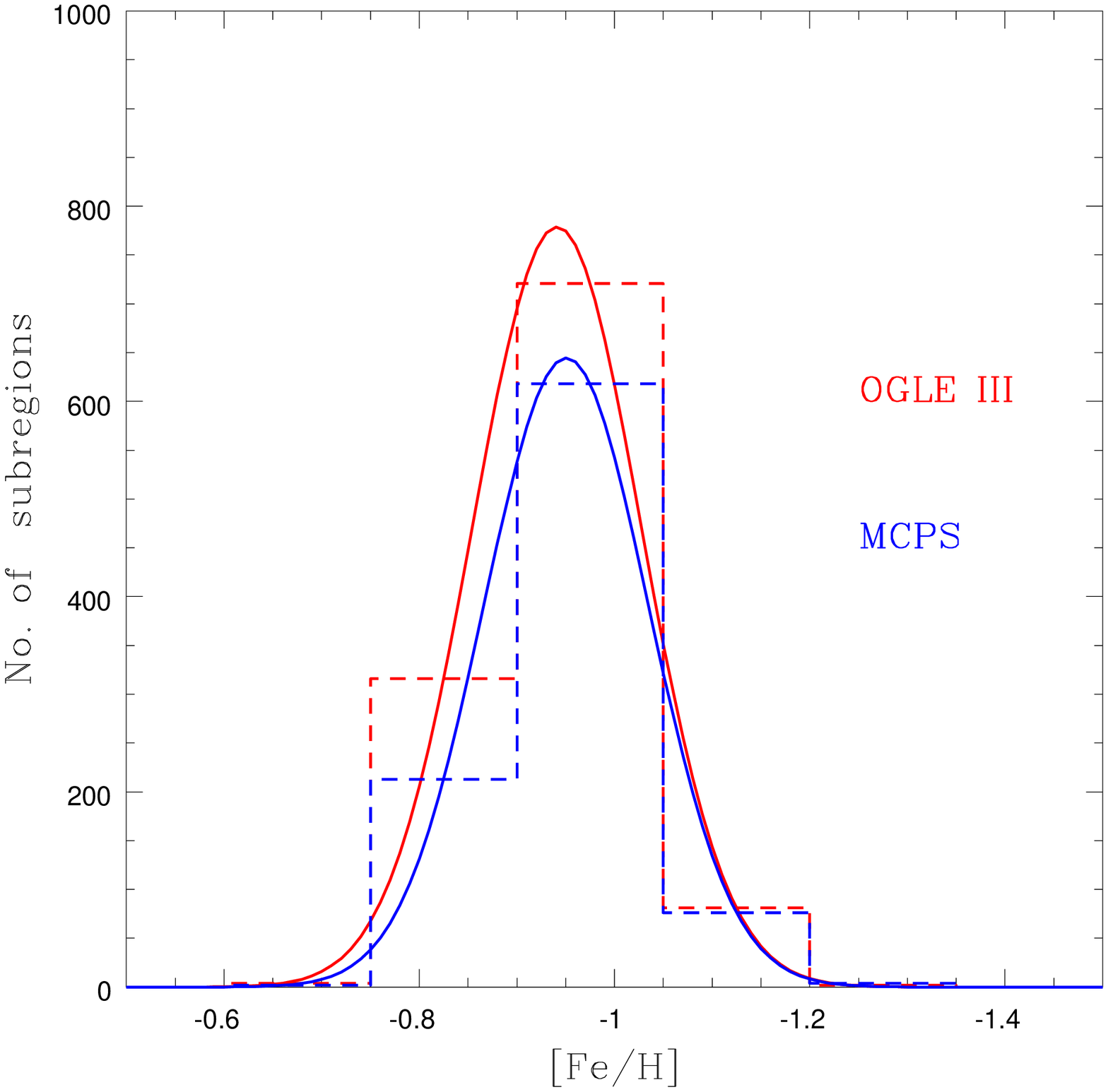}
\vskip 2cm
\caption{Histogram of metallicity (dashed lines) for the complete SMC fitted with a Gaussian function (solid lines), for cut-off criteria $IV$ of OGLE III (red) and MCPS data (blue).}
\label{fig:fig26}
\end{minipage}
\quad
\begin{minipage}[b]{0.45\linewidth}
\includegraphics[height=3.0in,width=3.0in]{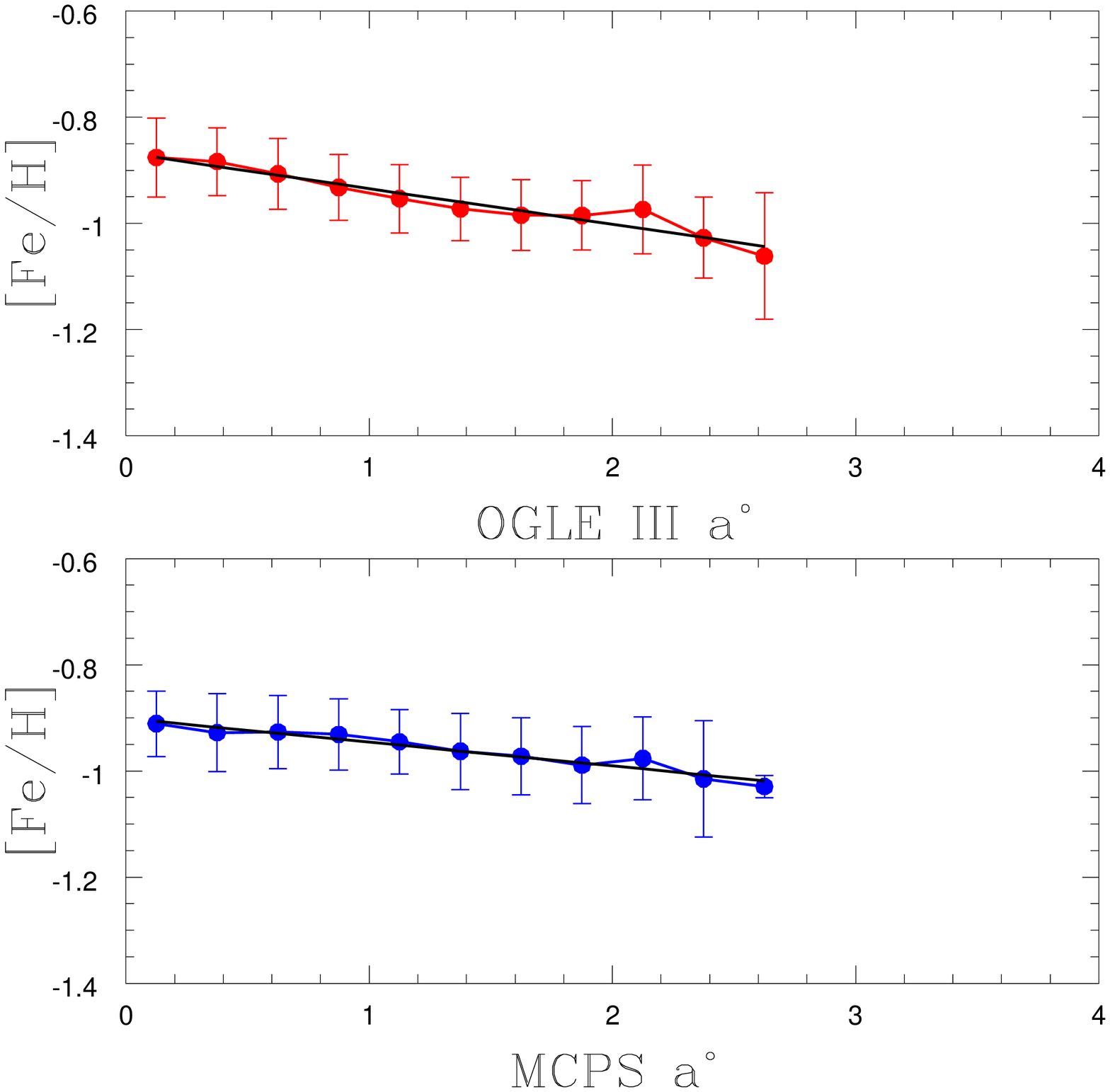}
\vskip 2cm
\caption{Variation of metallicity with semi-major axis (a) of the SMC for OGLE III (top panel) and MCPS (bottom panel) data, estimated for cut-off criteria $IV$. The MG estimated till $\sim$ 2.5${^\circ}$ is shown as black solid line for each panel.}
\label{fig:fig27}
\end{minipage}
\end{figure*} 

\subsubsection{Mean metallicity of the SMC}
In Figure \ref{fig:fig26}, we compare the distribution of metallicity for the entire SMC as estimated from the two data sets. It can be seen that in general the distributions are almost identical. The MCPS distribution has primary and secondary peaks at same metallicity values as OGLE III. The difference in the peak heights are due to the difference in number of subregions analysed for the two surveys. If we approximate the distribution to be Gaussian, then, for the OGLE III data, the peak of the distribution is estimated to be at [Fe/H] = $-$0.94 ($\sigma$[Fe/H] = 0.09). In the case of MCPS data, the peak of the distribution is found to be [Fe/H] = $-$0.95 ($\sigma$[Fe/H] = 0.08). Similar peak values and widths for both data sets suggests that the difference in area coverage does not have significant role in the derived metallicity distribution of the SMC. In other words, the Northern and Southern SMC seen in the MCPS do not have much variation in metallicity, which varies between $\approx$ $-$0.9--$-$1.0 dex. The range of metallicity within the SMC is found to be relatively less ($\approx$ $-$0.75 dex to $-$1.2) as compared to that found in the LMC in Paper I ($\approx$ $-$0.2 dex to $-$0.75 dex, Figure 36). An important point to be noted here is that, for the LMC in the MCPS analysis, the I magnitude of MCPS data were first transformed to that of OGLE III data, and then the re-estimated MCPS slopes were transformed to metallicity using the OGLE III slope-metallicity relation. However, in this work we have been able to estimate independent slope-metallicity relations for both the OGLE III and MCPS data sets respectively. Thus, there does not exist any effect in our metallicity estimations due to systematic differences between the OGLE III and MCPS filter systems. 

Our results for the mean metallicity of the SMC is in agreement with that of \cite{Carrera+2008AJ-CEH-SMC}, \cite{Parisi+2010AJfieldI, Parisi+2016AJfieldII} and \cite{Dobbie+2014MNRAS-papII}. \cite{Carrera+2008AJ-CEH-SMC} performed CaT spectroscopy to obtain metallicities of a sample of some 350 field red giants in 13 fields of size (8.85 $\times$ 8.85) sq. arcmin. distributed from $\sim$1$^{\circ}$ to 4$^{\circ}$ from the centre. They found a mean metallicity of [Fe/H] $\sim$ $-$1.0 dex in the innermost SMC fields ($<$ 2.5$^{\circ}$). Parisi et al.'s group have a series of work using CaT spectroscopy of RGs within the SMC field region \citep{Parisi+2010AJfieldI,Parisi+2016AJfieldII} and star clusters \citep{Parisi+2009AJclusI,Parisi+2015AJclusII}. \cite{Parisi+2010AJfieldI} estimated the metallicities of $\sim$ 360 red giant stars distributed in 15 SMC fields from the centre till about 8$^{\circ}$. The metallicity distribution of their whole sample had a mean value of [Fe/H] = $-$1.00 $\pm$ 0.02 dex, with a dispersion of 0.32 $\pm$ 0.01. However, the derived mean values for fields within 4$^{\circ}$ radius was about [Fe/H] = $-$0.99 $\pm$ 0.08 dex, whereas that beyond 4$^{\circ}$ was [Fe/H] = $-$1.02 $\pm$ 0.07 dex. Later, \cite{Parisi+2016AJfieldII} analysed a sample of 400 RGs within 15 fields, thus increasing their total sample of field stars to $\sim$ 750. The authors reported a median metallicity for this sample to be $-$0.97 $\pm$ 0.01 dex. The authors \cite{Parisi+2015AJclusII} using a combined sample of clusters from previous works yielded a high probability that the metallicity distribution for clusters is not uni-modal as the field stars but bi-modal, with peaks at $-$1.1  and $-$0.8 dex. Although we have used \cite{Dobbie+2014MNRAS-papII} to calibrate our metallicity map, we would like to state that our estimated mean metallicities using two large scale data sets are in accordance with their study. According to them, the metallicity distribution of their field star sample distributed within 5$^{\circ}$ of SMC centre, has a median metallicity of $-$0.99$\pm$0.01 dex which is similar to the mean obtained in this study. 

The mean metallicity of the SMC estimated by us is metal-rich as compared to previous studies using RR Lyraes by \citealt{Deb&Singh2010MNRAS, Haschke+2012AJ, Kapakos&Hatz2012MNRAS}, where the estimated mean is $\leq$ $-$1.50 dex. Given that we use RGB stars as indicators, the difference between our results is possibly related to the mean age difference between the RGBs and RR Lyraes. Due to mass and metallicity effects on RGB evolutionary rates, it is difficult for old, metal-poor populations to be the dominant contributor to the bulk average metallicity of red giants \citep{Manning&Cole2017}.

\subsubsection{Metallicity gradient of the SMC}

Next, we try to understand the metallicity distribution as a function of radius. To do so, the orientation of the SMC and projection effects must be addressed. The structure of the SMC is complex and less understood as compared to the LMC. According to theoretical and observational studies (\citealt{Bekki&Chiba2008ApJ, Zaritsky+2000ApJ, Harris&Zaritsky2006AJ,Evans&Howrath2008MNRAS}) the SMC is supposed to be a two-component system, where old and intermediate-age stars are distributed in a spheroidal or slightly ellipsoidal component, whereas, the young stars and gas are distributed in a disk. 

\cite{Smitha&Purni2012} studied the old RR Lyrae stars and the intermediate-age RC stars in the SMC and suggested that both these populations have a slightly ellipsoidal distribution and are located in a similar volume. The smooth ellipsoidal distribution of RR-Lyraes is also demostrated by recent works of \cite{JD+2017AcA} and \cite{Muraveva+2017arXiv} from analysis of large-area survey of the SMC in optical (OGLE IV)and near-infrared (VISTA-VMC) bands respectively. However, contrary to previous belief on the distribution of younger populations, \cite{JD+2016AcA} from their analysis Classical Cepheids using OGLE-IV data demonstrated the existence of a non-planar structure that can be described as an extended ellipsoid. \cite{Scowcroft+2016ApJ} using the mid-IR data of Cepheids from Spitzer, examined the three-dimensional structure of the SMC. The authors confirm that the Cepheid distribution does not just have a large line of sight depth, but is elongated from the north–east to the southwest, such that the southwestern side is up to 20 kpc more distant than the northeast. Also, very recently \cite{Ripepi+2017MNRAS} using near-infrared data (VISTA-VMC) show that the three-dimensional distribution of the Classical Cepheids is not planar, but heavily elongated for more than 25--30 kpc in the east/north-east towards south-west direction (approximately). Thus, the structure of the SMC remains a debated topic. It is also is believed that the SMC is markedly elongated along the line of sight (\citealt{Gardiner&Hawkins1991MNRAS, Haschke+2012AJ,Smitha&Purni2012,JD+2016AcA}), making projection effects important but the determination of true galactocentric distances difficult to ascertain.

We have estimated the radial metallicity gradient following the convention used by previous studies by \cite{Piatti+2007MNRASyoung, Parisi+2009AJclusI,Parisi+2010AJfieldI, Dobbie+2014MNRAS-papII}. This is shown in Figure \ref{fig:fig27}. For this, we considered an elliptical system whose major axis is positioned along the SMC bar (i.e along north-east -- south-west). We adopted this major axis to have an position angle of 55.3$^{\circ}$ east of north. If 'a' is the semi-major and 'b' is the semi-minor axis of the ellipse, we assumed the ratio a/b = 1.5 (as mentioned in \cite{Dobbie+2014MNRAS-papII} in accordance with \cite{Smitha&Purni2012}). For each star, we then estimated the value of 'a', that an ellipse would have if it were centred on the SMC, aligned with the bar, and one point of its trajectory coincided with star's position. This value of 'a' is used as a surrogate for the true galactocentric distance. We constructed a radial metallicity gradient, by assuming a bin width of 0.25$^{\circ}$, and  fitting a straight line using least square fit. The radial MG estimated for OGLE III is estimated to be: [Fe/H]= ($-$0.868$\pm$0.009) + a$\times$ ($-$0.067$\pm$0.006), $r$=0.97 ; for MCPS it is [Fe/H] = ($-$0.901$\pm$0.006) + a$\times$ ($-$0.045$\pm$0.004), $r$=0.97. The unit of the the MG is dex deg$^{-1}$. 

The estimated MG in the SMC is shallow and in accordance with that of previous studies by \cite{Carrera+2008AJ-CEH-SMC}, \cite{Parisi+2016AJfieldII} and \cite{Dobbie+2014MNRAS-papII}. We present a qualitative comparison between our results and some previous studies. Although \cite{Carrera+2008AJ-CEH-SMC} had mentioned the MG as one moves away from centre to larger galactocentric distances, their mean metallicities remain almost constant ($\sim$ $-$1 dex) till 2.5$^{\circ}$ from the centre and drops to metal poor values ($\sim$ $-$1.6 dex) only for two outer fields located between 2.5$^{\circ}$-4$^{\circ}$. The values for these two fields are also associated with relatively larger error. It needs a mention that Carrera et al's. results are based on fields distributed in the south, east and western regions, and none in the northern part of the SMC. The authors do not quantify the MG and do not consider an elliptical geometry. 

\cite{Parisi+2010AJfieldI} in their initial work claimed the non-existence of a MG. With their latest refined work, \cite{Parisi+2016AJfieldII} observed a MG in a larger sample of RGs. Their Figure 1 shows that their sample covers small pockets within the SMC except the North-West portion. While estimating the MG, they considered elliptical geometry (but with b/a = 1/2) and within the inner region of the galaxy (a $<$ 4$^{\circ}$) found a clear MG for the field stars of $-$0.08$\pm$0.02 dex deg$^{-1}$. On the other hand, in the outer part of the SMC (a $>$ 4$^{\circ}$) the authors report a positive MG, although the authors mark a word of caution that it requires detailed study to confirm this positive MG. We could not confirm this positive MG due the limitation in our spatial coverage. 

The  MG based on clusters has a different story. The combined cluster samples from \cite{Parisi+2015AJclusII, Parisi+2009AJclusI,DaCosta&Hatzidimitriou1998AJ, Glatt+2008aAJ, Glatt+2008bAJ} show that for a distance less than 4$^{\circ}$, they appear to be concentrated in two groups: one metal-rich group and the other metal-poor group relative to the field stars. The gradient of these two groups (metal-poor group: $−$0.03$\pm$0.05 dex deg$^{-1}$, with Y-intercept at $\approx$ $-$1.1 dex ; metal-rich group: $-$0.01$\pm$0.02 dex deg$^{-1}$, with Y-intercept at $\approx$ $-$0.75 dex ) are shallow and extremely dependent on the assumed vertex and on the edge definition made between the inner and the outer regions. Also, the statistical significance of these two potential gradients are difficult to assess. 

\cite{Dobbie+2014MNRAS-papII} considered similar geometry as this work, and estimated the MG to be $-$0.075 $\pm$0.011 dex deg$^{-1}$ within the inner 5 $^{\circ}$. The authors had covered a considerable space of about 37.5 sq. degrees about the SMC centre in all directions. However, since the authors dealt with individual stars, a lot of regions within the inner SMC were missed out for which we could estimate the mean metallicity. Overall, we see that the gradient estimated by \cite{Parisi+2016AJfieldII} and \cite{Dobbie+2014MNRAS-papII} seem a little steeper as compared to that estimated by us but are similar within the errors. The reasons could be the following: (1) We are estimating the mean metallicity of individual sub-regions and not stars, and then further smudging the mean metallicity while estimating the radial MG. (2) Our study is limited to $\sim$ 2.5 $^{\circ}$, whereas their work extends till 4 $^{\circ}$ - 5 $^{\circ}$. There can be an actual dip in mean metallicity beyond 2.5 $^{\circ}$ which can cause the MG to go down. In that case one needs to extend our technique using to the outer layers of the galaxy in order to understand the true nature of MG. (3) Our data sets have a more homogeneous distribution in space as compared to the above studies. \cite{Kapakos&Hatz2012MNRAS} estimated a very shallow, tentative MG ($-$0.013 $\pm$ 0.007 dex kpc$^{-1}$) in the de-projected plane of the SMC, by analysing V band light curves of 454 RR Lyraes using OGLE III data. However, as mentioned by the authors, the MG could be affected by selection effects and the confirmation of MG will require a detailed spectroscopic investigation.

Our results of MG are contrary to that put forward by \cite{Cioni2009A&Athemetallicity}, who using the AGBs estimated that the [Fe/H] has a constant value of $\sim$ $-$1.25$\pm$0.01 dex from centre up to $\sim$ 12 kpc. The authors estimated the gradient in the deprojected plane of the SMC considering its distance, Position Angle and inclination. The origin of this discrepancy with our study requires further investigation, and may be due to the difference in mean age between the general RGB field discussed here and AGB stars previously studied. The existence of MG is also in disagreement with \cite{Piatti2012MNRASage-metalSMC} who mentioned that there does not exists any metallicity or age gradient within the SMC, and with \cite{Haschke+2012AJ} and \cite{Deb+2015MNRAS} who do not detect any MG from the analysis RR Lyrae stars.

According to the studies of \cite{DaCosta&Hatzidimitriou1998AJ, Idiart+2007A&A, Carrera+2008AJ-CEH-SMC, Cignoni+2013ApJ} the most metal-rich stars within the SMC tend to be the youngest. \cite{Carrera+2008AJ-CEH-SMC} found a relationship between the MG and the age gradient, in the sense that the youngest stars are concentrated in the central metal-rich regions. \cite{Piatti2011MNRAS} from their study of SMC star clusters established an age-metallicity relation where they found a decrease from around [Fe/H] $\sim$ $-$0.5 to $-$1.0 dex for the 1-2 Gyr old populations down to typically [Fe/H] = $-$1.0 to $-$1.5 dex for the populations older than 5-6 Gyr. The author inferred that the younger population within the SMC is more concentrated towards the centre of the galaxy. 

\cite{Cignoni+2013ApJ} and Rubele et al. (2015) while analysing the star-formation history of the SMC field showed that the bar region and Wing region of the SMC have undergone recent star-formation and host younger population. Moreover, \cite{Cignoni+2013ApJ} mentioned that for similar age they do not observe any MG. \cite{Dobbie+2014MNRAS-papII} and \cite{Parisi+2016AJfieldII} assign their detection of MG to this increasing fraction of young stars in the inner regions of the SMC. From our analysis of a homogeneous data-set covering the inner SMC, we infer that the shallow but gradual MG that we detect in the inner SMC is possibly due to similar effect.

\begin{figure*}
\centering
\begin{minipage}[b]{0.5\linewidth}
\includegraphics[height=4.0in,width=4.0in]{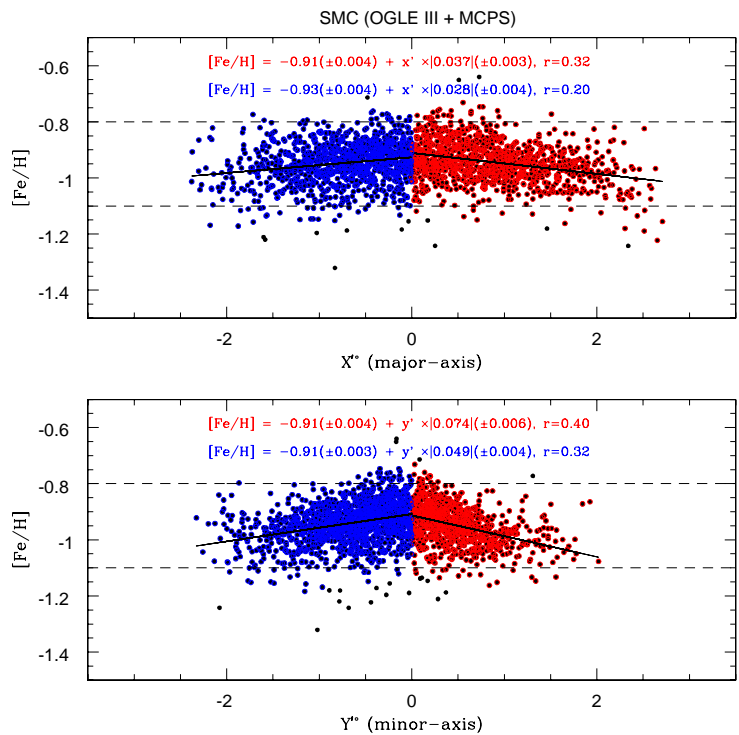}
\vskip 2cm
\caption{Estimated MG along the major (top panel) and minor (bottom panel) axis. The positive major axis is along south-west direction and the negative is along north-east. Whereas, the positive minor axis is along north-west direction and the negative is along south-east. The combined data points (OGLE III and MCPS - cut-off criteria $IV$) are plotted as black open circles in both the panels. The red open circles denote points after estimating MG with 3${\sigma}$ clipping along the positive major and minor axis. Whereas, the blue open circles denote points after estimating MG with 3${\sigma}$ clipping along the negative major and minor axis. The estimated MG, Y-intercept and $r$ are mentioned in both the panels corresponding to each case.}
\end{minipage}
\label{fig:fig28}
\end{figure*}
 
We plotted the combined metallicity distribution of OGLE III and MCPS data sets with respect to the major and minor axis of the SMC to check for any possible gradients along them. Figure 28 shows the variation of metallicity (black points) along the major and minor axis in the top and bottom panel respectively. The positive major axis is along south-west direction and the negative is along north-east. Whereas, the positive minor axis is along north-west direction and the negative is along south-east. The metallicity range in general is seen to be constrained between $-$0.80 dex and $-$1.10 dex for both the data set (with some deviations), which is $\pm$1$\sigma$ about the mean metallicity ($\approx$ $-$ 0.95 dex). These limits are shown by the dashed (black) lines parallel to the major/minor axis. There seems to exist a marginal metallicity gradient along the major axis, which is shallower relative to the gradient observed along the minor axis. 

To quantify the gradients, we made a linear least-squares fit with 3-$\sigma$ clipping along the positive (red points) and negative (blue points) major and minor axes. The estimates are labelled within the figures, where the metallicity gradients are denoted by their absolute values: 0.037 and 0.028 dex deg$^{-1}$ along positive and negative major-axis respectively; 0.074 and 0.049 dex deg$^{-1}$ along positive and negative minor-axis respectively. The MGs along the minor-axis are very similar to our estimated radial MG. The correlation-coefficient of the fits are also expressed by the absolute value of ($r$). The MGs estimated in this section indicate that it is not radially symmetric. The spread in metallicity values is more along the negative major and minor axis as compared to their positive counterpart. This is also reflected in $r$ being less along the negative major and minor axis as compared to their positive counterpart. In another words, in the north-east and south-east the metallicity is more dispersed spatially as compared to the north-west and south-west direction. This is possibly because the eastern side of the SMC is more perturbed due to its interaction with the LMC. 

\subsection{Spatial distribution of metal-rich regions within the SMC}

\begin{figure*}
\centering
\includegraphics[height=4.0in,width=7.0in]{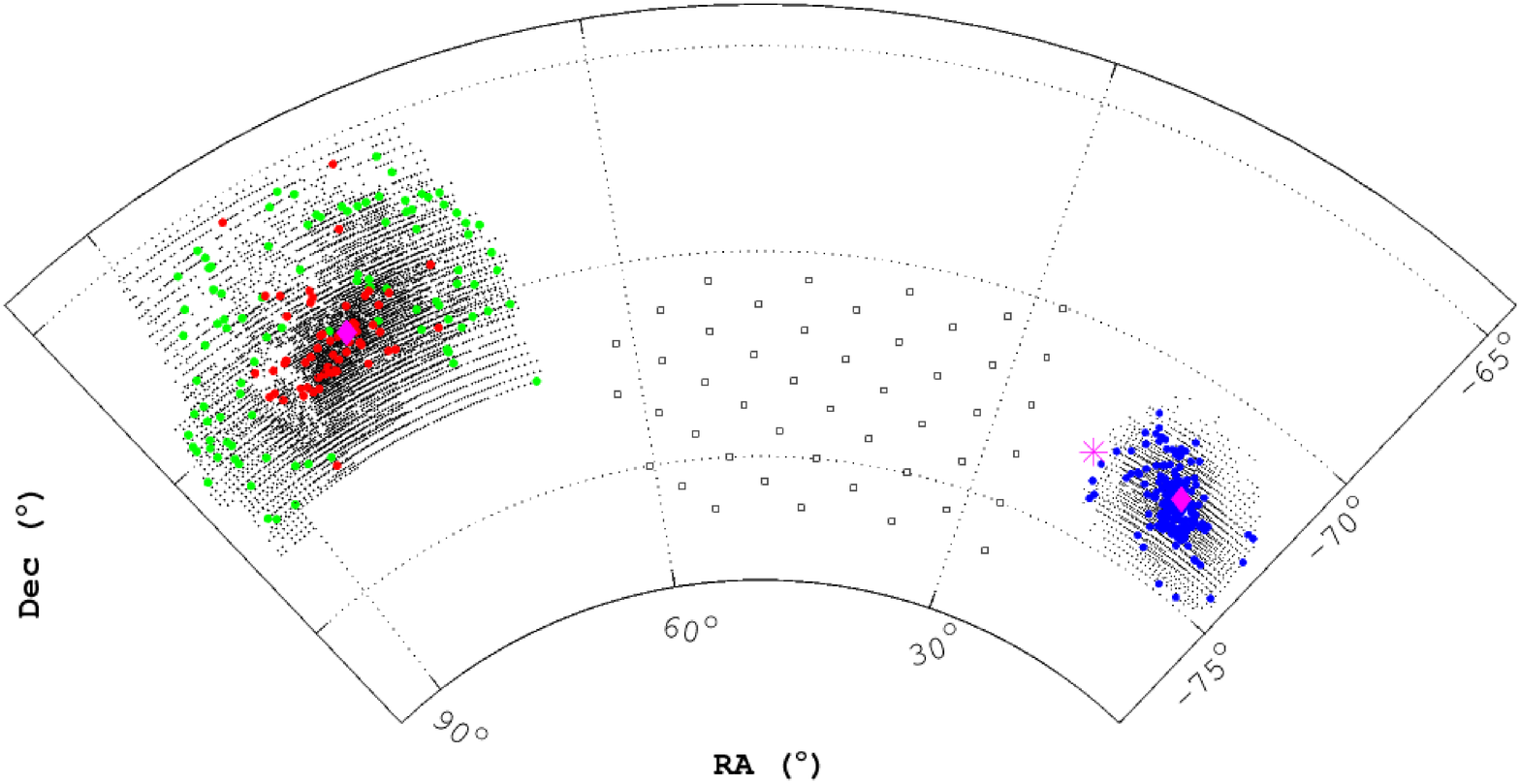}
\vskip 2cm
\caption{Combined metallicity map of the MCs using OGLE III and MCPS data for cut-off criteria $IV$. For the LMC, the metallicity outlier map from Figure 40 of Paper I is shown. The significantly metal rich ([Fe/H] range $-$0.05 to $-$0.15) and metal poor sub-regions ([Fe/H] range $-$0.65 to $-$0.95) within the LMC are shown in red and green colour respectively. For the SMC, the metal rich sub-regions ([Fe/H] range $-$0.60 to $-$0.85) within the galaxy are shown as blue points. The remaining points in the two galaxies are shown as black dots. The centre of the two galaxies are shown as magenta coloured diamonds. The centre of the SMC Wing is shown as magenta asterisk (Rubele et al. 2015). The main part of the MB is plotted as open squares \citep{Skowron+2014ApJ}, as observed by the OGLE IV survey region.
}
\label{fig:fig29}
\end{figure*}

In Figure \ref{fig:fig29} we present a spatial plot from the combined metallicity map of the LMC and the SMC using both OGLE III and MCPS data set (for cut-off criteria $IV$). The former is from Figure 40 of Paper I, where the outliers, i.e., significantly metal rich/poor regions with respect to the mean metallicity of the LMC ([Fe/H] $\approx$ $-$0.4 dex), are highlighted in red and green respectively. The significantly metal rich regions are distributed primarily along the LMC bar and have [Fe/H] $\geq$ $-$ 0.15 dex. Whereas, the metal poor counterparts have [Fe/H] in the range $-$0.65 dex to $-$0.95 dex. \cite{Olsen+2011ApJapopulation} in their work had reported a population of tidally accreted SMC stars in the outer regions of the LMC. However, we would like to remind the readers that there is no spatial correlation between the accreted population and the metal poor outliers as reasoned out in Paper I. To indicate the position of the MB we have shown its main portion as observed by the OGLE IV survey as analysed by \cite{Skowron+2014ApJ}. The centres of the LMC and the SMC are marked by magenta coloured diamonds. The centre of the SMC Eastern Wing is shown as magenta asterisk (Rubele et al. 2015). 

We find that the mean metallicity range near the central regions of the SMC ($\approx$ $-$0.6 to $-$0.85 dex) is very similar to the metal poor outlier in the LMC. We have highlighted these metal rich regions within the SMC in blue colour. We see that such regions are mostly spread around the centre along the SMC's major axis (i.e. along the bar). There also exists a scattered distribution of such regions in the SMC outskirts. These are more in the north-east region (in the direction of MB and Wing), when compared to the south-west. However, to connect this relatively large spread along the eastern direction to the evolution and interaction history of the SMC requires a detailed spectroscopic and kinematic analysis. The metal-rich regions in the south-west part spatially overlaps with the tentatively identified, metal-rich, kinematic counter-bridge structure in \cite{Dobbie+2014MNRAS-papI,Dobbie+2014MNRAS-papII}

We resorted to the same technique of RGB slope estimation for the LMC and the SMC. Also, we used results from CaT spectroscopy of RGs by the same group for both the MCs (\cite{Cole+2005AJspectroOfRGs}, \cite{Grocholski+2006AJCaIItriplet} for the LMC and \cite{Dobbie+2014MNRAS-papII} for the SMC) to calibrate the metallicity maps. Given that we expect only small systematic differences between the spectroscopic technique used for both the galaxies, the similarity in the metallicity range of the metal-rich regions within the SMC and the metal poor outliers within the LMC seems intriguing. On the other hand, given the values of mean metallicity of the LMC ($\approx$ $-$0.4 dex, Paper I) and the SMC ($\approx$ $-$0.95 dex, this work) as well as the metallicity range within these galaxies, this similarity is not surprising. We suggest a spectroscopic and kinematic analysis of the metal-poor outliers within the LMC and the metal-rich regions within the SMC to check for any similarities between these regions. Obtaining high-resolution, high-S/N spectra of stars within these areas would identify any commonalities or differences in the detailed distribution of abundances. The LMC metal-poor stars have a distinctive contribution from AGB-star winds to s-process elements and unusual elements, similar to alpha elements to the MW, but very different Cu abundances \cite{VanderSwaelmen+2013A&Achemiabun}. It would be of great interest to see if those traits are shared in the SMC at the same metallicity.


\section{Summary}
This paper presents an estimate of the average and radial variation of metallicity ([Fe/H]) in the SMC based on photometric data. The results can be summarised as follows:
\begin{enumerate}
\item We have successfully extended our technique of combining large scale photometric data and spectroscopic data developed in Paper I to estimate a metallicity map of the SMC. This again is a first of its kind high spatial resolution metallicity map for this particular galaxy derived using RGB stars from the OGLE III and MCPS data sets.
\item We estimate the RGB slope of several sub-regions in the SMC and convert the slope to metallicity using spectroscopic data of Red Giants in the field.
\item The average metallicity of the SMC is found to be $-$0.94 dex ($\sigma$[Fe/H] = 0.09) from OGLE III data and $-$0.95 dex ($\sigma$[Fe/H] = 0.08) from MCPS data, within a radius of 2.5$^{\circ}$. 
\item Using these large scale photometric data we confirm once and for all that there exists a MG within the inner SMC, which is in agreement with previous spectroscopic results. The estimated MG using both the data-sets is shallow and gradual from the SMC centre till a radius $\sim$ 2.5$^{\circ}$: from $-$0.045$\pm$0.004 dex deg$^{-1}$ for MCPS to $-$0.067$\pm$0.006 dex deg$^{-1}$ for OGLE III).
\item We find the MG within the SMC to be radially asymmetric. 
\item The metallicity range of metal-rich regions around the SMC centre ($\approx$ $-$0.6 to $-$0.85 dex) is similar to that of metal poor outliers located in the outskirts of the LMC. Such regions need to be studied in detail using spectroscopic studies to investigate the commonality.
\end{enumerate}
\section*{Acknowledgements}
This research was supported by the KASI-Yonsei Joint Research Program for all Frotiers of Astronomy and Space Science funded by the Korea Astronomy and Space Science Institute. The work is also partially supported by Basic Science Research Program through the National Research Foundation of Korea (NRF) funded by the Ministry of Education (NRF2016R1D1A1B01006608). Choudhury S. acknowledges Dr. Smitha Subramanian (Kavli Institute for Astronomy and Astrophysics, Peking University, Beijing, China) for suggestions during initial area binning of OGLE III and MCPS data, and also for critical reading of the original version of the manuscript and comments. The authors thank the OGLE and the MCPS team for making the data available in public domain. The authors also thank the anonymous referee for constructive suggestions, that helped improve the manuscript.

\bibliographystyle{mnras}
\bibliography{bibliography}

\bsp	
\label{lastpage}
\end{document}